\pdfoutput=1
\documentclass[journal]{IEEEtran}

\usepackage{cite}
\usepackage{stfloats}
\usepackage{amsmath, amssymb, accents, mathtools, amsthm} 
\usepackage{algorithm, algorithmicx, algpseudocode}
\usepackage{bbm}
\usepackage{mathtools}
\usepackage{graphics} 
\usepackage{epsfig} 
\usepackage{url}
\usepackage{multicol}
\usepackage{float}
\usepackage{epstopdf}
\DeclareMathOperator*{\argmin}{arg\,min}
\DeclareMathOperator*{\arginf}{arg\,inf}

\newcommand\stard[1]{\accentset{\star}{#1}}
\newfloat{twocolequfloat}{b}{zzz}
\floatname{twocolequfloat}{Equation}
\newtheorem{thm}{Theorem}
\newtheorem{lem}{Lemma}
\makeatletter
\newcommand*{\rom}[1]{\expandafter\@slowromancap\romannumeral #1@}
\makeatother

\begin{document}
\nocite{*}
\title{Optimal demand response policies for inertial thermal loads under stochastic renewable sources}

\author{

        Gaurav~Sharma$^1$,~\IEEEmembership{Student Member,~IEEE,}
        Le~Xie$^1$,~\IEEEmembership{Member,~IEEE,}
        and~P.~R.~Kumar$^1$,~\IEEEmembership{Fellow,~IEEE}
\thanks{ $^1$Department of Electrical and Computer Engineering, Texas A \& M University. Email:{\tt \{gash,le.xie,prk\}$@$tamu.edu.}} %
}
\maketitle



\begin{abstract}
In this paper, we consider the problem of preferentially utilizing intermittent  renewable power, such as wind, optimally to support thermal inertial loads in a microgrid environment. Thermal inertial loads can be programmed to preferentially consume from renewable sources. The flexibility in power consumption of inertial loads therefore can be used to absorb the fluctuations in intermittently available renewable power sources, and promote reduction of fossil fuel based costly non-renewable generation. Under a model which promotes renewable consumption by penalizing the non-renewable, but does not account for variations in the end-user requirements, the optimal solution leads to all the users' temperatures behave in a lockstep fashion, that is the power is allocated in such a fashion that all the temperatures are brought to a common value and they are kept the same after that point, resulting in synchronization among all the loads. In the first part, we showed that under a model which additionally penalizes the comfort range violation, the optimal solution is in-fact of desynchronization nature, where the temperatures are intentionally kept apart to avoid power surges resulting from simultaneous comfort violation from many loads. 

In the second part, the privacy of the end-user is additionally taken into account. We proposed an optimal demand response architecture where no information from the end-user is required to be transferred in-order to optimally co-ordinate their power consumption. We propose a simple threshold value based policy which is architecturally simple to implement, computationally in-expensive, and achieves optimal staggering among loads to smooth the variations in non-renewable power requirements. We show that the optimal solution for the threshold values is analytically commutable for a number of scenarios. We proposed a numerical approach to calculate the optimal solution for the scenarios where it is difficult to compute analytical solution. Lastly we propose a output feedback based adaptive heuristic approach to approximate the optimal solution for the scenario when sufficient information is not available to compute the optimal solution 
\end{abstract}



\section{Introduction}
The evolution of both the installed capacity and the consumption of renewable sources, such as wind and solar has been substantial over the last decade. 
Along with the higher utilization of renewable energy sources,  there is a significant interest in techniques to overcome the uncertainty in the generated power without losing the reliability to service the loads. In addition to generation side control, like automatic generation control, several demand response schemes are proposed to efficiently support variable generation sources \cite{short}, \cite{kundu1}, \cite{kundu2}, \cite{Callaway}, \cite{Callaway2}, \cite{ilic}.

In this paper, we propose to use thermal inertial loads to preferentially utilize the wind power than non-renewable power sources. Using such scheme lead to twofold benefits, first, it leads to reduced consumption of costly and polluting non-renewable generation, and second, the controllable thermal loads can absorb the temporary fluctuations in renewable sources. Some recent studies indicate that the thermal loads consumption takes above 60\% of the total energy consumption for an average consumer \cite{doe}. Therefore the thermal loads not only provide a substantial opportunity to be used as a controllable buffer, but also are cost effective as they do not require infrastructure changes, e.g. adding expensive energy storage units. Several interesting issues arises in using the thermal loads for such a demand response.  
 
The users of the thermostatic loads have a specified desired comfort range, within which they would like their temperature to lie. The change in comfort range may lead to temporary violation of comfort.  Except for the short period when comfort range is changed,  the demand response scheme is required to respect the imposed temperature limits. This leads to an important issue of how to design a policy that ensure such  adherence. Several optimization criteria can be used to, we discuss some of these formulations, and the nature of the resulting demand response that arise from using these criteria.

We consider a system in which both renewable power as well as costly non-renewable grid power is available from the main power grid. In order to achieve effective demand response to absorb fluctuating wind power, and utilize ``resource pooling" by collecting a large number of loads together we employ an architecture that consist of a central controller. We refer to such a controller as a load serving entity (LSE) (also ``load aggregaor" or simply an ``aggregator").

Under a model where we need to ensure that  temperature stays inside the comfort range at all times, during long periods when wind is not available, non-renewable power is required to supplement in order to avoid overheating. This leads to an important question of how to reduce the variations in the non-renewable power requirement from such a system. Low peak to average ratio is desirable as it leads to reduced consumption of costly operating reserves. 

For the scenario where there is no intermittent renewable generation, and only non-renewable power is used to cool a collection of thermal loads is studied by kundu et al \cite{kundu2}, their work shows that the collection which consumes the constant power results from a distribution where fraction of loads between two temperatures is proportional to the time taken to heat or cool between the two temperatures. This leads to a population which is evenly spaced in time and consumes a constant power lack of any variations.  The presence of stochastic wind power leads to an interesting complications. Wind is a common source available for all the loads, every load is cooled together whenever sufficient wind is available. This leads to a synchronization issue, that is the loads behave in a lockstep fashion. This is undesirable, because a sudden change in comfort range when wind power is not sufficient will lead to a hugh surge in total non-renewable power requirements to cool the loads. To address this issue we propose a scheme which explicitly models the comfort range variations, and showed that the under cost which penalizes the variations in non-renewable power consumption, it is optimal to stagger the loads. This ``symmetry breaking"  staggering hedges against the eventuality which may make all the load demand more non-renewable power due to  users' change in comfort range. This interestingly arises from a local concavity of the cost-to-go function in HJB equation.

One important issue at the load end is that of the privacy of individual users. Individuals do not want to reveal their thermal states to the aggregator. Revealing such information can be linked to other activities of a daily routines, working schedules etc \cite{pontryagin}. Thus we seek a solution where no instantaneos temperature measurement of load temperature is required to be communicated to the aggregator. This leads to an important issue of how an aggregator can influence the ``collective behavior" of the ensemble, i.e. their total power consumption without any information about individual temperatures. 

There are also issues related to communication system requirements. A system where agregator sends continous signal to each of the loads is not desirable, as it requires great communication bandwidth, and also raises concerns related to link reliability issues. Thus rather than seeking a centralized solution where the aggregator controls the cooling of each load, we seek a solution where each load controls its own cooling in a distributed fashion. 

Consider a collection of identical air-conditioners, each of them specify a comfort range $[\Theta_{min} , \Theta_{max}]$, within which its temperature should lie. Suppose renewable generation, say wind, is used to cool all homes whenever it is blowing. Non-renewable power is only used when the temperature hits $\Theta_{max}$. Under such a scheme, eventually all the loads will turn on and off at the same time. If due to some externality, loads change their $\Theta_{max}$ at the same time, this will result in a huge surge of non-renewable power, since all the loads will start consuming non-renewable power together upon hitting $\Theta_{max}$. Therefore such ``synchronized" behavior is undesirable. To avoid this, we proposed and analyzed a policy to stagger the loads within the comfort range by assigning each home a temperature at which it should start non-renewable power consumption. By selecting these thermostat setting parameters across the population of loads we can stagger the loads within its comfort range. Moreover the distribution of these parameters can be used to control the ``collective behavior" of the ensemble of loads. We show that such optimal staggering can be easily computed when number of homes is large. Moreover, this scheme allows each load's cooling to be controlled individually at its thermostat, and no information transfer is required from the loads to the aggregator.

The rest of the paper is organized like this. In section \rom{2} we discuss issues related to modeling of the wind power, comfort range setting and optimization schemes. We subsequently present a few results regarding the nature of the optimal demand response arising from the different optimization criteria, and give the underlying mathematical details explaining different nature of optimal demand demand response. In particular we show that the ``symmetry breaking" arises due to the  local concavity of the cost-to-go function of the HJB equation. In section \rom{3} we presented the privacy respecting architecture of demand response and give a method the explicit solution for load staggering under this architecture, we show that this optimal solution can be generalized for a variety of models. Finally, in section \rom{4} we give some simulation results for the privacy respecting architecture.  
\begin{figure}
\centering
\includegraphics[height=60mm,width=70mm]{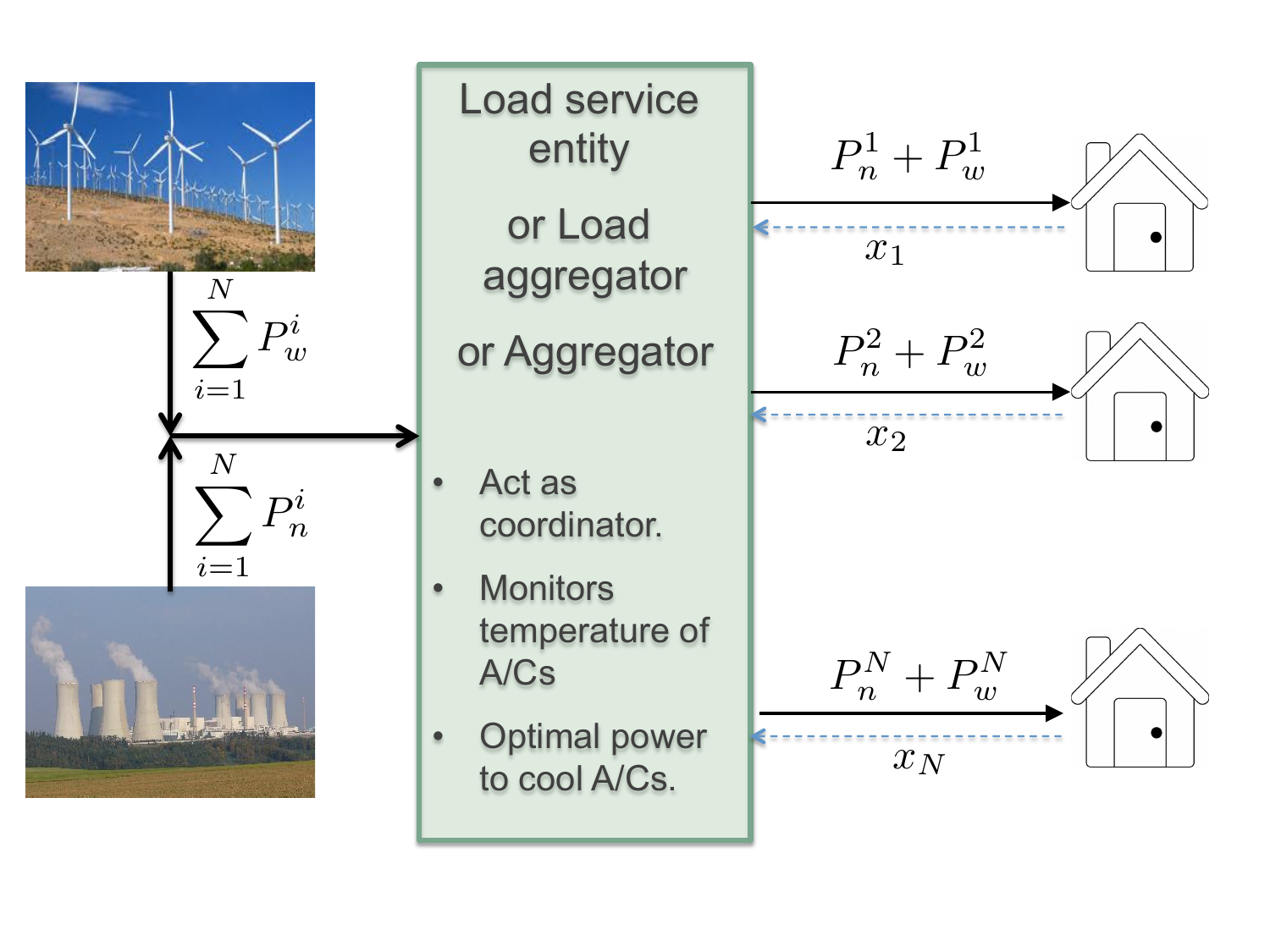}
\caption{In a general scheme, the Aggregator is aware of the loads' temperatures and wind process, and decides to optimal power components as a function of temperature and wind availability that should be used by a/c loads for cooling.}
\label{fig:nopriv}
\end{figure}

\section{Load dynamics, wind model and optimization criteria}
We consider an architecture consist of a central controlling agent. We refer to such central controller as load serving entity (LSE) (or aggregator). LSE has two functions, first it estimates the available wind power, and second  LSE controls the components of renewable power and non-renewable power to a collection of loads. Therefore LSE can control the cooling rate and the temperature of loads. We first assume that  LSE has full knowledge of the temperature state of all the loads, i.e. $\vec{x}$, and decides for each temperature $\vec{x}$ the amount of cooling and power each load should consume. This schemes is shown in figure \ref{fig:nopriv}. 

Air conditioner loads consumes power to cool themselves. The temperature of $i$-th load, $x_i(t)$ decreases with the total power consumed $P_i(t)$, where $P_i(t)=P_i^n(t)+P_i^w(t)$. Where $P_i^w(t)$ is the power from wind and $P_i^n(t)$ is the non-renewable power consumed by a load. We simplify the problem by relaxing the assumption that the loads cool at a fixed rate. We assume that the loads can cool at a fractional capacity, which can be achieved by switching between cooling 

The rate of change of temperature if a decreasing affine function in $P_i(t)$, namely $\dot{x}(t)= f_i(P_i)$. For illustrations, we often use  the dynamics $\dot{x}_i=h_i-P_i$. All the results will hold for a more general linear dynamics, for example the dynamics $\dot{x}_i= Ax_i+h_i-\alpha_iP_i$, proposed in several papers \cite{kundu1}, \cite{kundu2}, \cite{Callaway}. 

The comfort range for load-$i$ is assumed to be an interval $[\Theta_{min}^i,\Theta_{Max}^i]$. For simplification when considering identical loads we drop the index $i$, also we often shift the intervals, so that the lower limit is assumed to be fixed value $0$. To ensure that the comfort range adherence, we either 

With regard to modeling wind power as a stochastic process, there have  been several models proposed in the literature (refs). For simplicity we will assume a binary state wind model with two states, namely ``Blowing" and ``not Blowing" with mean holding times $1/q_1$, and $1/q_1$ respectively. When wind is in Blowing state there is some wind power $W$ is available for loads, whereas there is no power available in the not blowing state. We refer to this process as $\mathcal{M}(q_0,q_1)$ process.   

By a synchronizing policy we will mean a policy which reduces the difference in load temperature states by allocating more cooling power to a higher temperature TCL. On the contrary a de-synchronizing policy may occasionally increase the temperature difference by providing power to a lower temperature TCL and thus cooling it, while letting the higher temperature one increase.

We note that a de-synchronizing policy may be a state dependent that sometimes drives temperatures towards one another, and sometimes apart, i.e., it need not always strive to separate temperatures  all the time.

In what follows, we describe several alternative models and optimization criteria, and the nature of demand response that results from the combination of both.

\subsection{Contract violation probability (CVP) model:} 
In the presence of only renewable wind power,one model of a contract is to maintain load temperatures in a specified comfort range. This results in an objective of minimizing the probability of contract violation. Consider $N$ TCLs with temperatures denoted $(x_1,x_2,...,x_N)$, following the temperature dynamics $\vec{\dot{x}}=\vec{f}(\vec{P})$ as a function of supplied power $\vec{P}$. Suppose the contractual ranges are $\{[\Theta^i_{m},\Theta^i_{M}]\}^{N}_{i=1}$. Then we obtain the following optimization problem to minimize the contract violation probability:
\begin{align}
 \mbox{\bf Minimize }  & \sum \limits_{i=1}^N \mbox{ Prob } (  \{ x_i \notin [\Theta^i_{m},\Theta^i_{M}]\}) \nonumber\\
 \mbox{\bf Subject to } &  \frac{d\vec{x}}{dt} = \vec{f}(\vec{P}(t))\label{constr-dynamics} \\
  &\sum \limits_{i=1}^N P_i(t) \leq W(t) \label{constr-wind}\\
  &P_i(t) \geq 0 \mbox{ for } i=1,2,..N  \label{constr-nonneg}\\
  & W(t) \backsim \mathcal{M}(q_0,q_1) \label{constr-windproc}.
\end{align}
For this problem, since all loads outside the comfort zone contributes the same amount to cost, the loads which are nearer to the comfort zone are preferred by the optimization policy because they can be brought within the comfort range more quickly. Therefore some loads always remain cooler than others in this a model.

\subsection{Variance minimization model}
To correct unfairness in the last policy, a different model could be to reduce a penalty function which is square of the deviation above $\Theta_M$. The optimization problem is: 
\begin{align*}
 \mbox{\bf Minimize } & \int_0^T \bigg(\sum \limits_{i=1}^N  \mathbb{E} [(x_i(t)-\Theta^i_{m})^+]^2) | \vec{x}(0)]\bigg)\\
 \mbox{\bf Subject to } & \mbox{ Constraints (\ref{constr-dynamics}-\ref{constr-windproc})}.
\end{align*}
Due to the convex quadratic penalty, keeping the TCLs' temperatures apart from one another costs more than keeping their temperatures the same. The result is that it is in fact optimal to bring all TCLs to the same temperature and maintain them so. 
\subsection{Hard temperature threshold model} 
Notice that the preceding models does not guarantee any upper bound on the maximum temperature level. Such a hard temperature constraint cannot be met without a reliable power source. Therefore, we introduce a hard temperature upper bound in addition to the soft temperature comfort goal mentioned earlier. We also allow for but penalize the non-renewable power required to operate within the hard bound. This results in the following model
{
\begin{align}
 \mbox{\bf Min. } &\mathbb{E} \bigg[  \int_0^T \bigg(\sum \limits_{i=1}^N   [(x_i(t)-\Theta^i_{soft})^+]^{2}+ \gamma\bigg(\sum \limits_{i=1}^N P^n_i(t)\bigg)^2\bigg) \bigg] \nonumber\end{align}\begin{align}  
 \mbox{\bf Subject to } &  \frac{d\vec{x}}{dt} = \vec{f}(\vec{P}^w(t)+\vec{P}^g(t)) \label{constr1-dynamics} \\
  & x_i(t) \in [\Theta^i_{m},\Theta^i_{M}] \mbox{ for } i=1,2,...,N \label{constr1-temp}\\
  &\sum \limits_{i=1}^N P^w_i(t) \leq W(t) \label{constr1-wind} \\
  &P^w_i(t) \geq 0, P^n_i(t) \geq 0 \mbox{ for }i=1,2,...N \label{constr1-nonneg}\\
 & \mbox{and (\ref{constr-windproc})}\nonumber.
\end{align} }
In this model, at first sight one may speculate that keeping the TCLs apart hedges against their hitting the maximum temperature at the same time, thereby reducing the term in the cost function that is quadratic in the total power. However, it turns out that driving the TCLs to the same temperature and then maintaining them at equal temperature continues to be optimal. 
\subsection{Stochastic threshold variation (STV) model}
We now introduce the additional feature that there could be environmental, social or other extraneous events due to which the end user may change the set-point in a coordinated fashion. To capture this effect we will assume that there are two levels $\Theta_1$ and $ \Theta_2$, where $\Theta_1<\Theta_2$. All the TCLs switch between these levels at the same time instants according to a process $\Theta_M(t)\backsim \mathcal{M}(r_1,r_2)$, i.e. as a Markov process with mean holding times $1/r_1$ and $1/r_2$ in the two states $\Theta_1$ and $\Theta_2$.  Due to a sudden reduction in the set-point, a TCL that was previously within the desired temperature range $[0,\Theta_2]$ may suddenly be at a higher temperature than $\Theta_1$. When a TCL thereby violates  the threshold constraint we will require that it be  provided  grid power at the maximum possible level $M$ that the TCL can sustain, to cool it quickly: 
\begin{align*}
 \mbox{\bf Min. }& \int_0^T\mathbb{E} \bigg[\bigg(\sum \limits_{i=1}^N P^n_i(t)\bigg)^2 | \vec{x}(0) \bigg]\\
 \mbox{\bf s.t. } &  \mbox{ (\ref{constr1-dynamics},\ref{constr1-wind},\ref{constr1-nonneg}) and } \\
  & x_i(t) \in [0,\Theta_2] \mbox{ for } i=1,2,...,N\\
  & P^n_i= M \hspace{2mm}\{ \mbox{ if } x_i(t) > \Theta_{M}(t)\} \\
 &W(t) \backsim \mathcal{M}(q_0,q_1), \Theta_{M}(t) \backsim \mathcal{M}(r_{high},r_{low    }).
\end{align*}
We will show that with this model the optimal power allocation necessarily results in de-synchronization of the TCL temperature states. The optimal policy does not maintain all the TCLs at the same state. When there are TCLs above a certain level it is optimal to keep their temperatures different, to hedge against the future eventuality that the thermostats are switched down to $\Theta^1_{M}$. 
\subsection{HJB equation and nature of the optimal demand response}
For simplicity, we will consider a homogeneous population of TCLs, with identical temperature ranges, dynamics, heating, and user preference variations. The following results hold for the different models described in this section.
\begin{thm}
The optimal response in the CVP model with $\mathcal{M}(q_0,q_1)$ wind process and  linear dynamics is to provide all the wind power to the coolest TCL that is outside the temperature range. Therefore the optimal policy is {\it de-synchronizing}. 
\end{thm}    
\begin{thm}
Under the variance minimization model and $\mathcal{M}(q_0,q_1)$ wind process, the optimal control policy is of {\it synchronizing} nature, for any linear dynamics.
\end{thm}
\begin{thm}
Under the hard temperature threshold model with $\mathcal{M}(q_0,q_1)$ wind process and {constant} dynamics the optimal policy is of synchronizing nature.
\end{thm}
One can refer to \cite{kumar1} for proofs of the above results. Here we describe the origination of the {\it de-synchronizing} nature of the optimal policy for the {\it stochastic threshold variation} model.

Let $V^*_i(\vec{x},t)$ be the optimal cost-to-go from $\vec{x}$ at time $t$, in particular $$V^*(x)=\inf\{\mathbb{E}[\int_t^T (\sum_i P^n_i(t))^2dt|\vec{x}(t)=\vec{x}]\}$$  when wind state is $i$ (for $i=1$ when wind is {\it on}, $i=0$ when wind is {\it off}). In particular
The control variables are $\vec{P}^g \mbox{ and   } \vec{P}^w$, the allocated grid and wind power vectors to the TCLs. The state-space is $\mathcal{S}=[\Theta_m,\Theta_M]\times[\Theta_m,\Theta_M]\times\{0,1\}$. Let $\widehat{\mathcal{A}}(\vec{x},i) \subset \mathbb{R}_+^4$ be the set of state dependent {\it admissible} actions.
\begin{align*}
\widehat{\mathcal{A}}(\vec{x},i) = \{(\vec{P}^g,&\vec{P}^w): P^w_1+P^w_2\leq W \mbox{ if (Wind $i$=1)},\\
& \mbox{ if (Wind $i$=0) } P^w_1=P^w_2=0, \\
& \mbox{ if }(x_j(t)=0) \mbox{ then } f(P^n_j+P^w_j) \geq 0, \\
& \mbox{ if }(x_j(t)=\Theta_M) \mbox{ then } f(P^n_j+P^w_j), \leq 0 \\
& \mbox{ if }(x_j(t)>\Theta_M) \mbox{ then } P^n_j=M\}.
\end{align*}  
The Hamilton-Jacobi-Bellman equation is,
{
\begin{align}
&\inf_{\substack{(\vec{P}^g,\vec{P}^w) \in \widehat{\mathcal{A}}(\vec{x},i) }}
[(\mathbf{1}^{T}\vec{P}^g)^2 + \nabla^T V^*_i(\vec{x},t)\vec f(\vec{P}^g+\vec{P}^w)]\nonumber\\
& -q_iV^*_i(\vec{x},t)+q_iV^*_{!i}(\vec{x},t)+\frac{\partial {V}_i^*}{\partial t}(\vec{x},t) =0 \hspace{2mm} \mbox{ For } i=0,1.\label{eqn:hjb} 
\end{align}
}
For homogeneous loads with the state dynamics $f(P)=h-P$ (here $h$=heating due to ambient temperature when $P=0$), the infimum term for control separates. So for $x_i \in (0,\Theta_M)$, the optimal wind and grid power allocations are given by
\begin{align}
\vec{P}^{*g}(\vec{x},i)&=\arginf_{\vec{P}^g\geq 0}
\left((P^n_1+P^n_2)^2-\frac{\partial V^*_i}{\partial x_1}P^n_1 -\frac{\partial V^*_i}{\partial x_2}P^n_2 \right)\label{arg2-1}\\
\vec{P}^{*w}(\vec{x})&=\arginf_{(P^w_1+P^w_2=W)}\left(-\frac{\partial V^*_1}{\partial x_1}P^w_1 -\frac{\partial V^*_1}{\partial x_2}P^w_2\right).\label{arg2-3}
\end{align}  
whose solution is given by 
\begin{align}
(\stard{P}^w_1,\stard{P}^w_2)&= 
\begin{cases} \mbox{(W,0)} & \mbox{ if } \frac{\partial V^*_1}{\partial x_1} >\frac{\partial V^*_1}{\partial x_2} \\
\mbox{(0,W)}  & \mbox{ if }\frac{\partial V^*_1}{\partial x_1} <\frac{\partial V^*_1}{\partial x_2}
\end{cases}.\label{eqn:soln2w}
\end{align} \begin{align} 
(\stard{P}^g_1(\vec{x},i),&\stard{P}^g_2(\vec{x},i)) = \\ 
&\begin{cases}
(\frac{1}{2}\frac{\partial V^*_i}{\partial x_1}(\vec{x}),0) & \mbox{ if } \frac{\partial V^*_i}{\partial x_1} >\frac{\partial V^*_i}{\partial x_2}\\
(0,\frac{1}{2}\frac{\partial V^*_i}{\partial x_2}(\vec{x})) & \mbox{ if } \frac{\partial V^*_i}{\partial x_1} <\frac{\partial V^*_i}{\partial x_2}\\
(\frac{1}{2}\frac{\partial V^*_i}{\partial x_1}(\vec{x}), \frac{1}{2}\frac{\partial V^*_i}{\partial x_2}(\vec{x})) & \mbox{ if } \frac{\partial V^*_i}{\partial x_1} =\frac{\partial V^*_i}{\partial x_2}.
\end{cases} \label{eqn:soln2} 
\end{align} 
It is easy to see that $V^*(\vec{x})$ component-wise increasing function,  however the transition in behavior happens based on the concavity or convexity of the cost-to-go function $V^*$. In particular, when $V^*(\vec{x})$ is convex, for $x_1>x_2$ we get $(P_1^w,P_1^n)=(W,\frac{1}{2}\frac{\partial V^*_1}{\partial x_1}(\vec{x}),0)$ and $(P_2^w,P_2^n)=(0,0)$. So power is only allocated to the hotter load to cool, while the cooler house heats up. This results in their temperature difference to decreases, till the point their temperature becomes equal. After this their temperature is kept the same. 

The situation reverses when cost-to-go function is concave, that is, for $x_1>x_2$, we get $(P_1^w,P_1^n)=(0,0)$ and $(P_2^w,P_2^n)=(W,\frac{1}{2}\frac{\partial V^*_2}{\partial x_2}(\vec{x}))$. Thus the cooler load gets all the power to cool itself further. This results in futher seperation the temperature of both the loads.     

Based on the nature of this optimal de-synchronization, we can propose a heuristic policy which coola the coolest load when temperature of the collective loads become sufficiently high, other than the cases when renewable power is sufficient to cool all the loads.

\begin{figure}
\centering
\includegraphics[height=60mm,width=70mm]{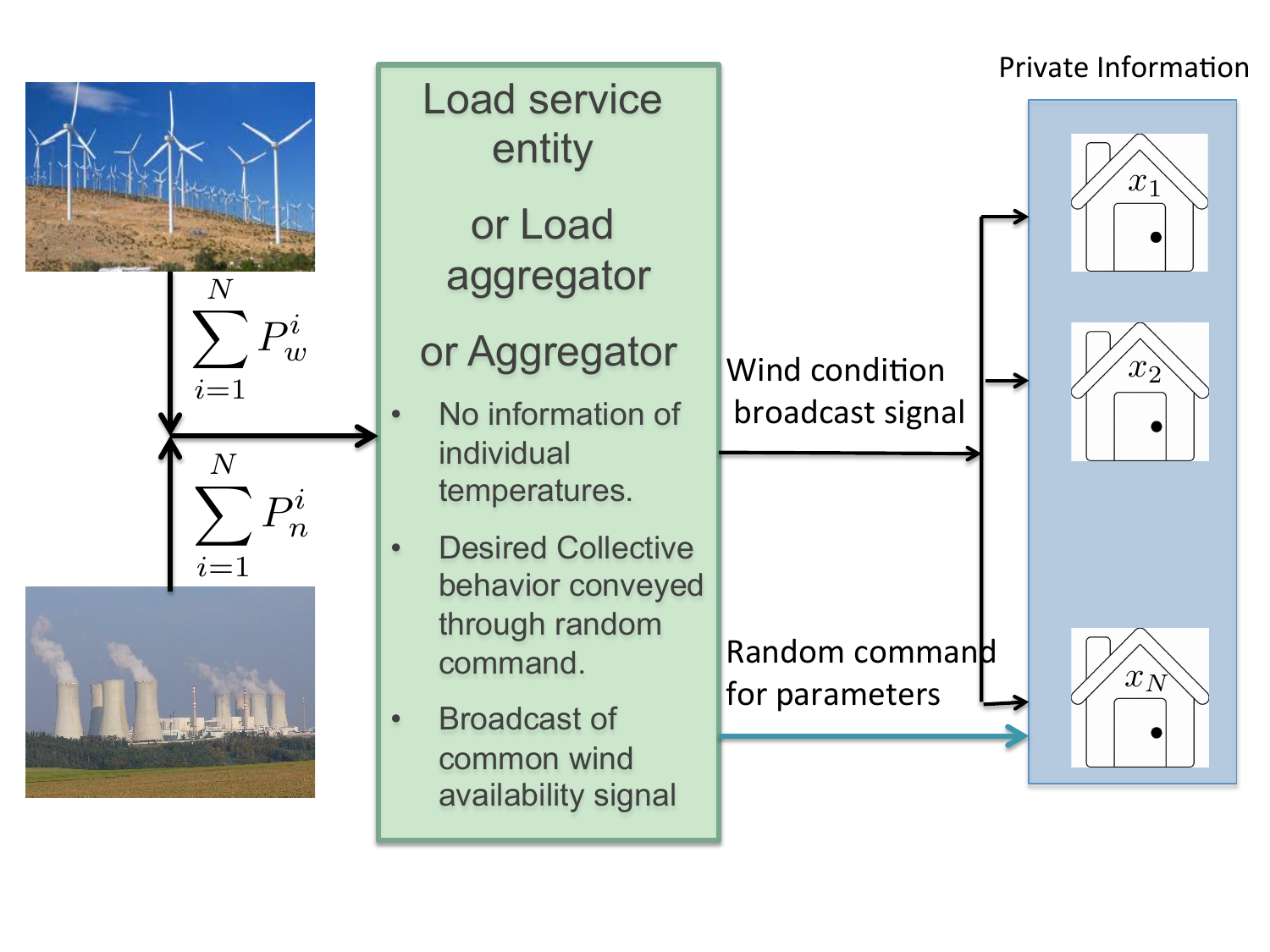}
\caption{In a privacy respecting demand response model, the states of the loads are not revealed to the aggregator. The aggregator controls the aggregate power consumption behavior by setting the parameters randomly and broadcasts the wind availability to all the loads to control the ensemble of air-conditioner loads. }
\label{fig:priv}
\end{figure}

\section{Privacy respecting architecture using a threshold policy}

There are issues with previous policy, first for a large number of houses it becomes difficult to solve the optimization problem. The HJB Equation \ref{eqn:hjb} is nonlinear, therefore can not be solved directly, also by using numerical methods the complexity increases exponentially with  number of houses. Second, since control action is a function of temperature, the houses need to send their temperature information back to the LSE. This is both expensive with respect to communication requirements, and is intrusive due to the revelation of the state variable to the controller. 

Motivated by these shortcomings, in this section we will introduce a set-point policy. We will illustrate that the optimal solution for this set-point policy is easy to compute. In-fact, under some condition the policy turns out to be an explicit function when the population size becomes infinite.

The aggregator under threshold policy provides a ``wind availability" signal to the loads, which can also be regarded as akin to a ``price signal". The loads choose their own comfort level set-points. Under this threshold policy, each load $i$, uses the wind upon availability to cool itself, unless it is already at the lowest temperature allowed in comfort range. 

When wind is not available the temperature of the loads rises. Consider a collection of identical thermal inertial loads, each of which under a simple model follow the following temperature dynamics, $$\dot{x}_i=h-P$$ when load is supplied with power $P$. Where $h$ denotes the ambient heating effect. We further assume the maximum cooling power is $h+c$, resulting in maximum cooling rate $c$. 

Further, we assume the identical loads  have a common desired comfort range $[0,\Theta_M(t)]$, within which they would like their temperature to lie. For simplicity, we denote the lower limit of comfort range by zero. To prevent the sudden demand increase resulting from simultaneous comfort range, we assume that the upper comfort setting is variable. Where we observe, that whenever this comfort setting change happens, temperatures of some load may exceed their upper comfort level. Under such situation, we use the non-renewable power to cool at a maximum cooling rate $c$, to restore the user-provided comfort range specifications at the earliest.  

We illustrate the thresold policy with a simple case, we assume that upper comfort level $\Theta_M(t)$ is a piecewise constant Markov process with two levels $\Theta_1$ and $\Theta_2$ with mean holding times $\frac{1}{r_1}$ and $\frac{1}{r_2}$, respectively. Further, let wind is also a two-state Markov process with states ``Blowing" and ``Not Blowing", with respective mean holding times $\frac{1}{q_1}$ and $\frac{1}{q_0}$. 

\begin{figure}
\centering
\includegraphics[height=45mm,width=80mm]{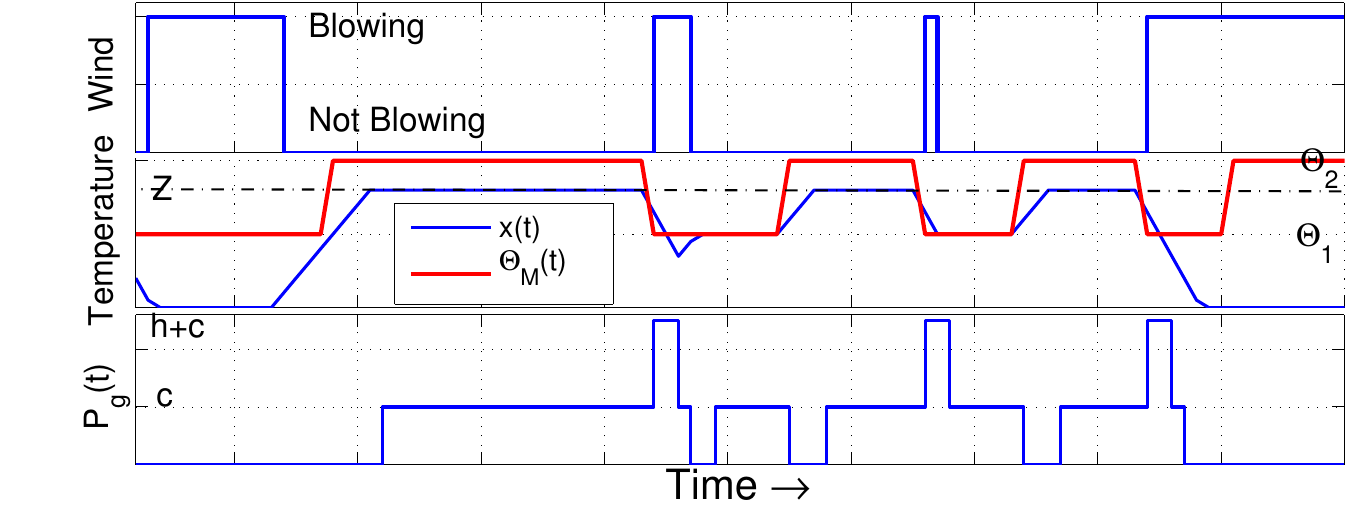}
\caption{In the $Z$-policy, a load consumes wind power whenever available to maintain its comfort range. When wind power is not available, it is allowed to heat till its temperature heats up to a value $z$ or the maximum temperature  in the comfort range, then it consume just enough power to maintain that temperature level.}
\label{fig:zpolicy}
\end{figure}

When wind is in ``Blowing" state, each load use wind to cool at maximum rate, unless they are already in lowest possible state, i.e. at $x=0$. At $x=0$, loads use just enough wind power to prevent over-cooling. Therefore we have 
\begin{align*}
\dot{x}(t)=\begin{cases}
-c & \mbox{When $x>0$ and wind is blowing}\\
 0 & \mbox{When x=0 and wind is blowing}
\end{cases}
\end{align*} 
When wind is in ``Not Blowing" state, and the load is inside comfort range $(0,\Theta_M(t))$, loads heat up at the rate  $h$ when they don't use non-renewable power. The loads use the non-renewable power whenever the loads are outside comfort-range $[0,\Theta_M(t)]$, and cool themselves at the maximum rate $c$ in order to return to comfort range quickly. In addition, in a threshold policy,  each load $i$ is assigned with a threshold temperature $Z_i$. Each load is allowed to heat till it hits either their upper comfort level $\Theta_M(t)$, or the threshold temperature level $Z_i$, upon hitting either, the loads will draw just enough non-renewable power to maintain itself at the temperature. This is shown in Figure \ref{fig:zpolicy}. 

Therefore the temperature follows 
\begin{align*}
\dot{x}_i(t)=\begin{cases}
0 &\mbox{if $x_i(t)=\min(Z_i,\Theta_M(t))$ \& wind not blowing}\\
h &\mbox{if $x_i(t)<\min(Z_i,\Theta_M(t))$ \& wind not blowing}\\
-c &\mbox{if $x_i(t)>\Theta_M(t)$ unconditionally}
\end{cases}
\end{align*}   

Let $P_w(t)$ and $P_n(t)$ denote the wind and non-renewable power, for the policy described above, we have 
\begin{align*}
P_w(t)=\begin{cases}
h+c & \mbox{when $x(t)>0$}\\
h   & \mbox{when $x(t)=0$}
\end{cases},
\end{align*}
and, 
\begin{align*}
P_n(t)=\begin{cases*}
h+c &\mbox{When $x(t)>\Theta_M(t)$ and wind not blowing}\\
h   &\mbox{When $x(t)= \min(Z,\Theta_M(t))$}
\end{cases*}
\end{align*}

For a collection of $N$ loads, with  the $i$-th load using the threshold $Z_i$, where $0\leq Z_i\leq \Theta_2$. Let $\mathbf{Z}=(Z_1,Z_2,...,Z_N)$, and the overall threshold policy as ``$\mathbf{Z}$-policy". Let $x_i(t)$ be the temperature of load-$i$ at time $t$, and $P_{n,i}(t)$ be the non-renewable power it draws. When the user changes its comfort level, it may occasionally violate the temperature range constraint, we penalize such violations by $[(x_i(t)-\Theta_M(t))^+]^2$ (where $x^+:=\max(x,0)$), and consider the average cost of such discomfort $\lim_{T\to \infty}\frac{1}{T}\int_0^T\sum_{i=1}^N [(x_i(t)-\Theta_M(t))^+]^2dt$.     

We also require the total non-renewable power drawn $\sum_{i=1}^N P_{n,i}(t)$ to be as constant as possible, we penalize the variations in total non-renewable power drawn by imposing a cost quadratic cost function $\lim_{T\to \infty} \frac{1}{T}\int_0^T (\sum_{i=1}^N P_{n,i}(t))^2$.

We therefor consider the problem of optimally selecting $\mathbf{Z}=(Z_1,Z_2,...,Z_N)$ to optimize the overall cost  $$\lim_{T \to \infty}\int_0^T \frac{1}{T} \{(\sum_{i=1}^N P_{g,i}(t))^2+\gamma^{(N)}\sum_{i=1}^N[(x_i(t)-\Theta_M(t))^+]^2dt\},$$ where, $\gamma^{(N)}$ trades off discomfort against less variations in total non-renewable power consumption. 

Although we have considered a square cost in both power consumption and discomfort, any other cost function of total non-renewable power $C_{power}(\sum_{i=1}^N P_{n,i})$, and discomfort to the customer $C_{discomfort}(x_i(t))$ would result in similar analysis.

\subsection{Evaluating distribution under threshold policy}
\label{sec:distr}
In order to evaluate the long-term average terms in the cost function, as a first step we need to evaluate the probability distribution under. For two wind levels (``Blowing" and ``Not blowing"), and  two upper comfort limits ($\Theta_1, \Theta$), let $p^z_{ij}(x)$ denote the probability density functions, after employing the z-policy for $x\in (0,\Theta_1)\cup (\Theta_1,\Theta_2)$, where $i=0,1$ denote wind ``Not blowing", and ``Blowing" respectively, and $j=0,1$ denote $\Theta_M$ equal to $\Theta_1$, and $\Theta_2$ respectively. Since, under the threshold policy, the load will remain at $0,\Theta_1,z$ each for a non-zero fraction of time, there will be a probability mass at each of these temperatures. Denote by   
$\delta_{0,j}^z$, the probability mass at $x=0$ and comfort setting $j$ as described above, also let $\delta_{\Theta_1}^z$, and $\delta_z^z$ denote the probability mass function at temperatures $\Theta_1$ and $z$ respectively. 

One may notice that each of the four probability mass functions $\delta_{0,0}^z, \delta_{0,1}^z, \delta_{\Theta_1}^z$, and $\delta_z^z$, are defined such that they can only be associated with a unique wind and comfort level settings, and as a result of this it is easier to evaluate the relation between density function $p_{ij}$ and probability mass at the boundary. One can also notice that the total probability mass at $x=0$ is $\delta_0^z=\delta_{0,0}^z+\delta_{0,1}^z$.

\begin{lem}
The density functions $p^z(x):=[p_{00}^z,p_{01}^z,p_{10}^z,p_{11}^z]$, and probability mass $\delta_{0,0}^z, \delta_{0,1}^z, \delta_{\Theta_1}^z$, and $\delta_z^z$  under a z-policy for binary state wind and binary comfort settings, is given by the following linear system;
\begin{align}
&D(x)\frac{d}{dx}p^z(x)=Qp^z(x) \label{dynamics}\\
&Q_1\delta^z_{\Theta_1}=D(\Theta_1-)p^z(\Theta_1-)-D(\Theta_1+)p^z(\Theta_1+)\label{boundary1}\\
&Q_2\delta^z_z=D(\Theta_2-)p^z(\Theta_2-)\label{boundary2}\\
&Q_3\delta^z_{0,1}+Q_4\delta^z_{0,1}=D(0+)p^z(0+)\label{boundary3}\\
&\int_0^z 1^T p^z(x)dx+\delta_{0,0}^z+\delta_{0,1}^z+\delta_{\Theta_1}^z+\delta_z^z=1 \label{normalization}
\end{align}

Where $D(x)$ is a diagonal matrix representing the dynamics of the stantes , Where $D_{ii}(x)=\frac{dx}{dt}$ where $i=1,2,3,4$ denote the states $ij=00,01,10,11$ respectively. Therefore, for the description of z-policy in last section, we have  $(D_{22},D_{33},D_{44})(x)=(h,-c,-c)$ and $D_{11}(x)=\begin{cases} h & \mbox{for $x<\Theta_1$} \\ -c &\mbox{for $x>\Theta_1$} \end{cases}$ 
$Q$ denotes the generator matrix for Markov process whose column are $Q_1,Q_2,Q_3,Q_4$ denote the transition from states $ij=00,01,10,11$ respectively. 
\end{lem} 
\begin{lem}
\label{lem:conserv}
Density function $p^z(x)$ satisfies the conservation law $1^TD(x)p^z(x)=0 ,\forall x\in [0,\Theta_2]$.
\end{lem}

\begin{proof}
In Equation (\ref{dynamics}), $Q$ is the generator matrix of a Markov process, thus $1^TQ=0$, which yields $\frac{d}{dx}1^TD(x)p^z(x)=0$. Also we have $1^TD(\Theta_i)p^z(\Theta_i)=0$, at each of the boundary points (Equations (\ref{boundary1}-\ref{boundary3})). So we have $1^TD(x)p^z(x)$ conserved at each temperature $x$.  
\end{proof}
The above method can also be generalized to more states of wind levels and comfort setting. In particular when there are $\mathcal{W}$, wind states and $\mathcal{C}$ comfort setting levels below temperature threshold z, thus $\mathcal{W}\times\mathcal{C}$ number of states. Differential equation (\ref{dynamics}) will still be the same, and will represent $\mathcal{C}$ linear differential equation involving $\mathcal{C}$ vectors of $\mathcal{W}\mathcal{C}$ variables, resulting in a total of $\mathcal{C}^2\mathcal{W}$ variables. 

To solve for these we will need to evaluate the boundary conditions, Let $\delta_{0,i}^z$ denotes the probability mass at $x=0$, under wind condition $i, i\in\{1,2,...\mathcal{W}-1\} $ and comfort settings $\Theta_j, j\in\{1,2,...,\mathcal{C}\}$, and let $\delta_{\Theta_j}^z$ denote the probability mass at $\Theta_j, j\in\{1,2,...,\mathcal{C}\}$.  This lead to additional $\mathcal{W}\mathcal{C}$ variables. 

To solve for these $\mathcal{C}^2\mathcal{W}+\mathcal{W}\mathcal{C}$ variable we have the $\mathcal{C}+1$ relations relating delta variables at $\{0,\Theta_1,...,\Theta_\mathcal{C}\}$ to each of $\mathcal{C}\mathcal{W}$ states, thus a total of $(\mathcal{C}+1)\mathcal{W}\mathcal{C}=\mathcal{C}^2\mathcal{W}+\mathcal{W}\mathcal{C}$ relations.  We will still need the normalization equation, since their is a dependency due to conservation (Lemma \ref{lem:conserv}). Thus we can obtain the distribution for arbitrary wind levels and comfort level setting. 

In fact, the same formulation exits for any other dynamics too. The dynamics matrix $D(x)$ is dynamics dependent, and captures the effect of different dynamics. In particular, when the cooling dynamics is $\dot{x}=Ax+b$, can be solved in a similar fashion. 

\begin{figure}[!h]
\centering
\includegraphics[height=30mm,width=70mm]{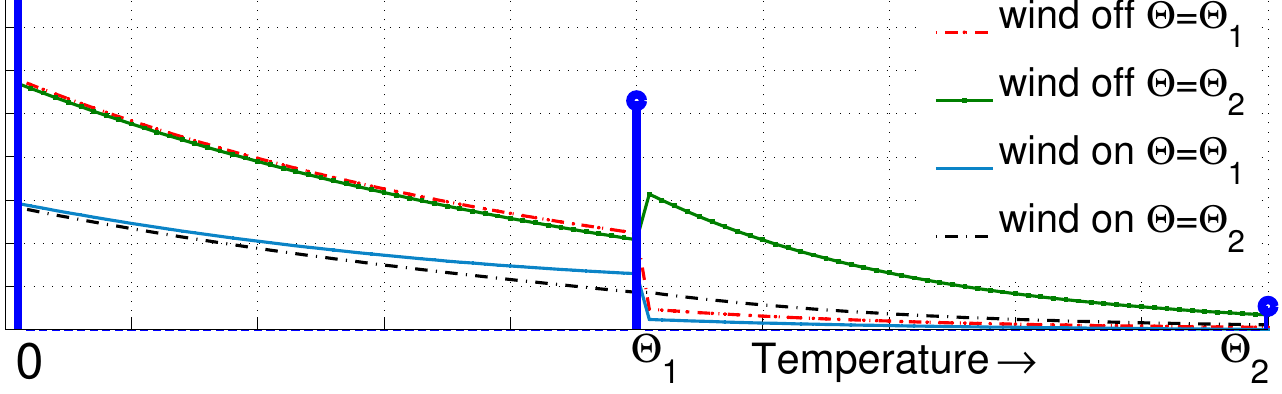}
\caption{A particular distribution of the set-point policy for values $(z,\Theta_1,\Theta_2,q_0,q_1,r_0,r_1,c,h )=(100,100,50,.04,.04,.02,.02,1.1,1)$. The probability
distribution consists of three point masses and four probability distributions which are continuous everywhere except at $\Theta_1$}
\label{fig:distribution}
\end{figure}

\subsection{Optimization for finite loads case under Z-policy}
\begin{figure}[!h]
\centering
\includegraphics[height=35 mm,width=90mm]{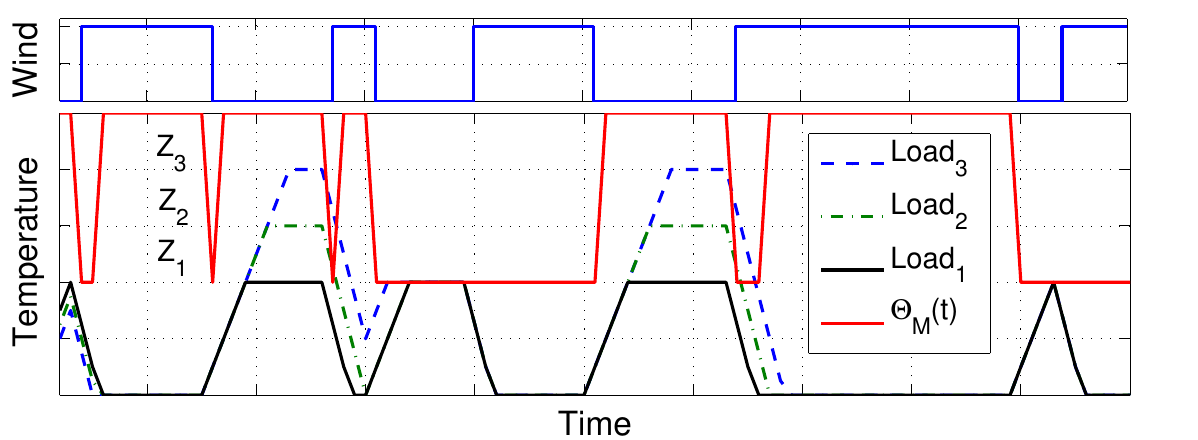}
\caption{When the wind is blowing, loads cool at the maximum rate, subject to not going below temperature 0, load temperatures do not rise above  minimum of $Z$ and $\Theta_M(t)$}
\label{fig:zthree}
\end{figure}

Figure \ref{fig:zthree} shows typical temperature trajectories for the case when $N=3$, and ${\bf Z}=(60,70,80)$ with $\Theta_2=100$. Since we are considering the average cost problem, we can assume without loss of generality that all loads start with the same initial temperature. In order to calculate the average cost, we need the distribution of $\sum_{i=1}^N P_{g,i}$, which entails knowing the joint probability distribution of $\{ P_{g,i}\}_{i=1}^N$. More generally, as we consider a large number of loads, which we do in the sequel, we need the joint distribution of a large number of loads' grid power draws. In order to determine this, an important property that we will exploit is that of stochastic domination. For any two loads $i$ and $j$  with $Z_i<Z_j$  we have $x_i(t) \leq x_j(t)$ for all $t$, regardless of the wind process realization. An important consequence is that whenever the $j$-th load hits its upper temperature limit $Z_i$, the $i$-th load has already hit its upper temperature limit too. Using $\omega$ to denote a sample point in the probability space, we have the following inclusion of events:
\begin{align}
 \{ X_j(t,\omega)=Z_j \} \subset \{ X_i(t,\omega)=Z_i \}, \hspace{5mm}\mbox{ if } Z_i<Z_j\label{observ}.
\end{align}
Now we can calculate the expected cost of a $Z$-policy. First consider just one TCL. The fossil fuel power drawn is $h$ when $\Theta(t)=\Theta_2$ and the load temperature is at $Z$, while it is $h+c$ when $X(t)>\Theta(t)$. The cost due to discomfort is  $\gamma^{(1)}[(x(t)-\Theta(t))^+]^2$, For any event $A$, denoting $\mathbb{P}^i(A):=\mathbb{P}(A\cap \{ \Theta(t)=\Theta_i \})$ for $i=1,2$. The cost of the $Z$-policy is 
\begin{align*}
&C^{(1)}( Z)= [(h+c)^2 \mathbb{P}^1(X>\Theta_1)+h^2\mathbb{P}^1(X=\Theta_1)]+\\
&\hspace{5mm}h^2\mathbb{P}^2(X=Z)+\gamma^{(1)} \int_0^z [(x-\Theta_1)^+]^2(p^z_{00}(x)+p^z_{10}(x))dx.
\end{align*}
For brevity let us denote the second term by $\Phi(z)$,
\begin{align*}
&\Phi(z):= \int_0^z [(x-\Theta_1)^+]^2(p^z_{00}+p^z_{10})(x)dx.
\end{align*} 
Now we consider the case of two loads. In this case we need to consider the \emph{total} dispatchable fossil fuel generation drawn by the two loads. Thus we need the \emph{joint probability distribution} of the two loads. Their marginals will not suffice, unlike the case of just one load. It is here that we exploit the above stochastic dominance which provides information on the joint-distribution of the loads. When  $(\Theta_M(t)=\Theta_2)$ with $Z_1\leq Z_2$,  the total fossil fuel power is $2h$ when $X_2=Z_2$ because $X_2=Z_2 \Rightarrow X_1=Z_1$. Also, when $X_2<Z_2$ but $X_1=Z_1$, the  fossil fuel power is $h$. Similarly when $\Theta_M(t)=\Theta_1$, total  fossil fuel power is $2(h+c)$ when $X_1>\Theta_1$, $(h+c)+h$ when $X_1=\Theta_1$ but $X_2>\Theta_2$, and $2h$ when $X_2=\Theta_2$. So the total cost $C^{(2)}$ with relative weight $\gamma^{(2)}>0$ is, 
\begin{align*}
C^{(2)}&({\bf Z})=(2h)^2\mathbb{P}^2(X_2=Z_2) +h^2\mathbb{P}^2(X_2<Z_2 \cap X_1=Z_1)\\
      &+(2h)^2\mathbb{P}^1(X_2=\Theta_1)+(2(h+c))^2 \mathbb{P}^1(X_1>\Theta_1)\\
      &+((h+c)+c)^2\mathbb{P}^1(X_2>\Theta_1\cap X_1=\Theta_1)\\
      &+\gamma^{(2)}(\Phi(Z_1)+\Phi(Z_2)).
\end{align*}
Also, since,  $\{X_2=Z_2\}\subset\{X_1=Z_1\}$,  we have $\mathbb{P}(\{X_1=Z_1\}) =\mathbb{P}(\{X_1=Z_1\}\cap \{X_2<Z_2\})+\mathbb{P}^2(\{X_2=Z_2\})$, i.e., 
\begin{align}
\mathbb{P}^2(\{X_1=Z_1\}\cap \{X_2<Z_2\}) &=\delta^{Z_1}_{Z_1}-\delta^{Z_2}_{Z_2}. \label{probdiff1}
\end{align}
When $\Theta(t)=\Theta_1$, by a similar argument we obtain,
\begin{align}
\mathbb{P}^1(\{X_1=\Theta_1\}\cap \{X_2>\Theta_2\}) &=\delta^{Z_1}_{\Theta_1}-\delta^{Z_2}_{\Theta_1}. \label{probdiff2}
\end{align}
This analysis extends to the case of $N$ loads. For any choice $(Z_1,Z_2,...,Z_N)$ of the set points with $Z_1\leq Z_2 .... \leq Z_N$, we can evaluate the value of the average. It is  
{
\begin{align*}
&C^{(N)}= \gamma^{(N)}\sum_{k=1}^N\Phi(Z_i)+N^2[h^2(\mathbb{P}^2(X_1=Z_1) +\\
&\hspace{20mm}\mathbb{P}^1(X_N=\Theta_1))+(h+c)^2\mathbb{P}^1(X_1>\Theta_1)]+\\
 &\sum_{k=1}^{N-1} \Big[((N-k)h)^2 \mathbb{P}^2(X_{k+1} = Z_{k+1} \cap X_{k} < Z_{k})+ \\
 &\big((N-k)(h+c)+(kh)\big)^2 \mathbb{P}^1(X_k=\Theta_1 \cap X_{k+1} >\Theta_1)\Big]
\end{align*}}
Now we turn to an important issue concerning the choice of the scaling parameter $\gamma^{(N)}$. $\int_0^T(\sum_{i=1}^N P_{g,i})^2dt$ grows like $\Omega(N^2)$, but  $\int_0^T \sum_{i=1}^N[(x_i(t)-\Theta(t))^+]^2 dt$ grows like $\Omega(N)$. We therefore scale $\gamma^{(N)}$ as $\gamma^{(N)}=\gamma.N$. Let $\widehat{C}^{(N)}:=\frac{C^{(N)}}{N^2}$  denote the normalized cost. It evaluates to
{
\begin{align}
& \widehat{C}^{(N)}= \frac{\gamma}{N} \sum_{k=1}^N\Phi(Z_i)+h^2\{\mathbb{P}^2(X_1=Z_1) +\\
&\hspace{20mm}\mathbb{P}^1(X_N=\Theta_1)\}+(h+c)^2\mathbb{P}^1(X_1>\Theta_1)+\nonumber\\
 &\sum_{k=1}^{N-1} \Big[\big(1-\frac{k}{N}\big)^2h^2 \mathbb{P}^2(X_{k+1} = Z_{k+1} \cap X_{k} < Z_{k})+\nonumber \\
 & \Big(\big(1-\frac{k}{N}\big)(h+c)+\big(\frac{k}{N}h\big)\Big)^2 \mathbb{P}^1(X_k=\Theta_1 \cap X_{k+1} >\Theta_1)\Big]
\label{Ncase}
\end{align}}
We thereby arrive at the following optimization problem for the case of a finite number $N$ of TCLs:
\begin{align}
\mbox{Minimize}\hspace{3mm} &\widehat{C}^{(N)}({\bf Z}) \nonumber\\
\mbox{s.t.}\hspace{3mm} & Z_i\leq Z_{i+1} \mbox{ for } i=1,2,...,N-1, \nonumber\\
& Z_1 \geq 0, Z_N \leq \Theta_2. \label{finitecase}
\end{align} 
This problem is intractable because of the large number of constraints. Motivated by this we next examine its infinite load population limit.

\subsection{Continuum limit for binary wind and comfort setting model}
We now consider the infinite population limit where there is a  continuum of loads. Let $u(z)$ denote the fraction of loads with set points no more than $z$. Let $\cal{U}$ denote the  space of piecewise continuous increasing positive functions on $[0,1]$, noting that $u \in \mathcal{U}$. The resulting cost, following a similar analysis to (\ref{Ncase}), is

\begin{align*}
C^{[0,1]}&(u)=\gamma\int_0^{\Theta_2}\Phi(z)u'(z)dz+ h^2(\delta^{\Theta_2}_{\Theta_2}+ \delta^{\Theta_1}_{\Theta_1})  \\
&+\int_0^{\Theta_2}[(hu)^2(z)\mathbb{P}^2(\{X_z=z\} \cap \{X_{z+dz}<z+dz\}))\\
&+ (hu(z)+(h+c)(1-u(z)))^2 \times \\
& \hspace{10mm}\mathbb{P}^1(\{X_z=\Theta_1\}\cap \{X_{z+dz}>\Theta_1\})].
\end{align*} 
The first term is the cost of violation of upper temperature limit, while the second term is  that due to variability in total dispatchable fossil fuel generation.
 Note that from  (\ref{probdiff1}),(\ref{probdiff2})  $\mathbb{P}^2(\{X_z=z\} \cap \{X_{z+dz}<z+dz\})= -\frac{d\delta_z^{z}}{dz}dz$  and $\mathbb{P}^2( \{X_z=\Theta^1\} \cap \{X_{z+dz}> \Theta^1\})= -\frac{d\delta_{\Theta_1}^{z}}{dz}dz$. 

Define  $D_1(z):=-\frac{d\delta_z^z}{dz}, D_2(z):=-\frac{d\delta_{\Theta_1}^z}{dz}$. The above can be simplified  (as in \cite{kumar2}) to
{
\begin{align*}
C^{[0,1]}(u)= &h^2(\delta^{\Theta_2}_{\Theta_2}+\delta^{\Theta_1}_{\Theta_1})+\bigg[\int_0^{\Theta_2} [\gamma u'(z)\Phi(z)+\\
&(hu(z))^2D_1(z)+((h+c)-cu(z))^2D_2(z)]dz\bigg].
\end{align*}}
Ignoring the first term that does not deoend on $u$, we obtain the following Calculus of Variations problem:

\begin{align}
\hspace{-2mm}
\mbox{Min}\hspace{.2mm} J[u]=&\int_0^{\Theta_2} \big[(hu(z))^2D_1+(h+c-cu(z))^2D_2+ \nonumber\\ &\hspace{10mm}\gamma\Phi(z)u'(z)\big]dz\label{cost}\\
\mbox{s.t.}\hspace{5mm}& u \in \mathcal{U}, u(0)=0, u(\Theta_2)=1. \label{cond1} 
\end{align} 

If one informally uses the Euler-Lagrange equation, one obtains the resulting solution $ u_{EL}(z)=\frac{\gamma\Phi'(z)+2c(c+h)D_2(z)}{2(h^2D_1(z)+c^2D_2(z))}$. This however need  not  be a positive increasing function or satisfy the boundary condition $u_{EL}(0)=0$, $u_{EL}(\Theta_2)=1$. Figure shows a particular solution where $u_{EL}\notin \mathcal{U}$ and $u_{EL}(\Theta_2)\neq 1$.

\begin{figure}[!h]
\centering
\includegraphics[height=20mm,width=80mm]{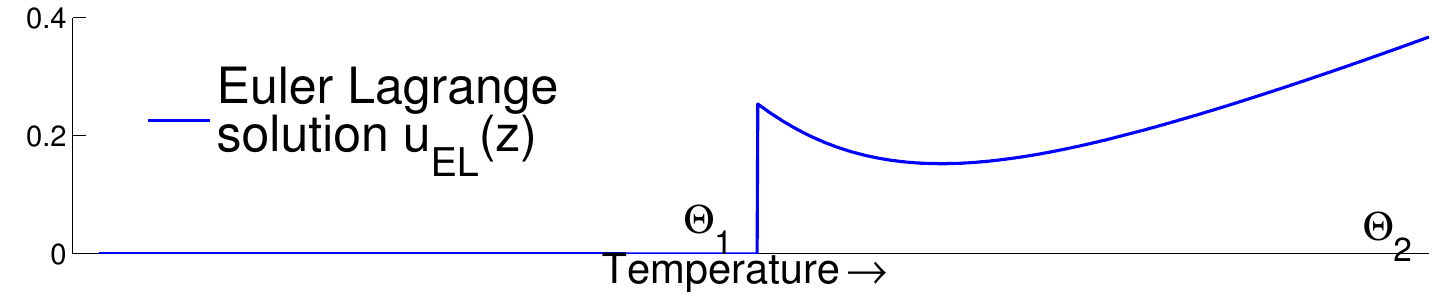}
\caption{The Euler Lagrange solution obtained for the values used to find distribution in Figure \ref{fig:distribution}.}
\label{uel}
\end{figure} 

Therefore the Euler Lagrange solution is  not  admissible for all the scenarios. We consider next how to solve this problem using Pontryagin's Minimum Principle.

\subsection{Evaluation of optimal solution using minimum principle}
\label{sec:pmp}
We employ the theory of optimal control to solve the optimization problem. Assume that state $u(z)$ follows the dynamic system $\dot{u}(z)=f(u,v,z)$  for  any admissible control $v(z)\in \bf{V}$, and $z \in [0, \Theta_2]$. Consider the cost function is  $J[u]= \int_{0}^{\Theta_2}  g(u,v,z)dz$.
 
Defining $h(u,v,\lambda,z)=g(u,v,z)+\lambda(z)f(u,v,z)$, the optimal control $v^*(z)$ which takes the state $u(z)$ from $u(0)=0$ to $u(\Theta_2)=1$, satisfies the following two conditions \cite{pontryagin}:
\begin{enumerate}
\item $\dot{\lambda}(z)= - \frac{\partial g}{\partial u}$ 
\item $v^*(z)\in \mathbf{V}$ is the pointwise minimizer of  $h(u,v,\lambda,z)$. 
\end{enumerate}  

Using Theorem 3.36 from \cite{folland}, since $u(z)\in \mathcal{U}$ has bounded variation, and $\Phi(z)$ is continuous, we have 
\begin{align}
\int_0^{\Theta_2} \Phi(z) u'(z)dz &= \Phi(\Theta_2) -\int_0^{\Theta_2} \Phi'(z) u(z)dz. \label{byparts}
\end{align}
We use (\ref{cost}),(\ref{byparts}) to collect terms depending on $u(z)$, to rewrite the cost in the following equivalent way:
{
\begin{align*}
J'[u]&= \int_0^{\tiny\Theta_2}  u^2(z)w(z)-u(z)(\gamma \Phi'(z)+2c(c+h)D_2(z))dz \\
&= \int_0^{\Theta_2} \bigg[u(z)-u_{EL}(z)\bigg]^2w(z)  \\
&\hspace{20mm}- \frac{(\gamma \Phi'(z)+2c(c+1)D_2(z))^2}{4(h^2D_1(z)+c^2D_2(z))} dz.
\end{align*}}
Where  $w(z):=(h^2D_1(z)+c^2D_2(z))>0$. Thus, we will focus on optimizing the cost functional  $J[u]=\int_0^{\Theta_2} w(z)[u(z)-u_{EL}(z)]^2dt$, $u(0)=0$ and $u(\Theta_2)=1$.  Wlog, we can take $u_{EL}(z) \in [0,1]$ (otherwise replace $u_{EL}$ by $\max(0,\min(1,u_{EL}))$).


We first consider the Euler Lagrange solution $u_{EL}(z)$ obtained in Section \rom{4}. In the notations above, we identify $\mathbf{V}=\mathbb{R}$, $f(u,v,z)=v^2$ and $g(u,v,z) = w(z)[u(z)-u_{EL}(z)]^2$.

The optimizers $v^*,\lambda^*$ satisfy, 
\begin{align}
&\dot{\lambda}^*(z)=-2(u(z)-u_{EL}(z))w(z) \label{lamdot} \\
&\dot{u}(z)=v^{*2}(z) \label{xdot}\\
&v^*(t)=\argmin_{v \in R} \bigg[w(z)(u(z)-u_{EL}(z))^2+\lambda^*(z) v^2(z)\bigg] \label{ustar}
\end{align}  

One can rewrite (\ref{ustar}) as:
{
\begin{align*}
v^*(z)=\begin{cases}
0 &\mbox{if $\lambda^*(z) > 0$} \\
-\infty &\mbox{if $\lambda^*(z) <0$ } \\
\mbox{arbitrary} &\mbox{if $\lambda^*(z)=0$ .}
\end{cases}
\end{align*}}

One can observe that $\lambda^*(z)$ can-not be strictly negative, as this leads to unbounded $u(z)$ from (\ref{xdot}), and further results in infinite cost $J$. This also suggests that jumps in $u(z)$ are possible only at those points where $\lambda^*(z)$ is zero.

Also from (\ref{lamdot}), we observe that $\lambda^*(z)$ remains constant only on the points where $u(z)=u_{EL}(z)$, it  increases when $u(z)<u_{EL}(z)$, and decreases when $u(z)>u_{EL}(z)$. From (\ref{xdot}), (\ref{ustar}) we observe that if $\lambda^*(z)>0$, then $v^*(z)$ is zero, leading to a constant $u(z)$. So, $\lambda^*(z)>0$ implies $u(z)=\mbox{constant}$. Next, since $u(z)$ has to be increasing from (\ref{xdot}), we conclude that $\lambda^*(z)$ can-not be constant on the points where $u_{EL}(z)$ is decreasing, however $\lambda^*(z)$ can be constant when $u_{EL}(z)$ is increasing.

\begin{figure}[!h]
\centering
\includegraphics[height=30 mm,width=85mm]{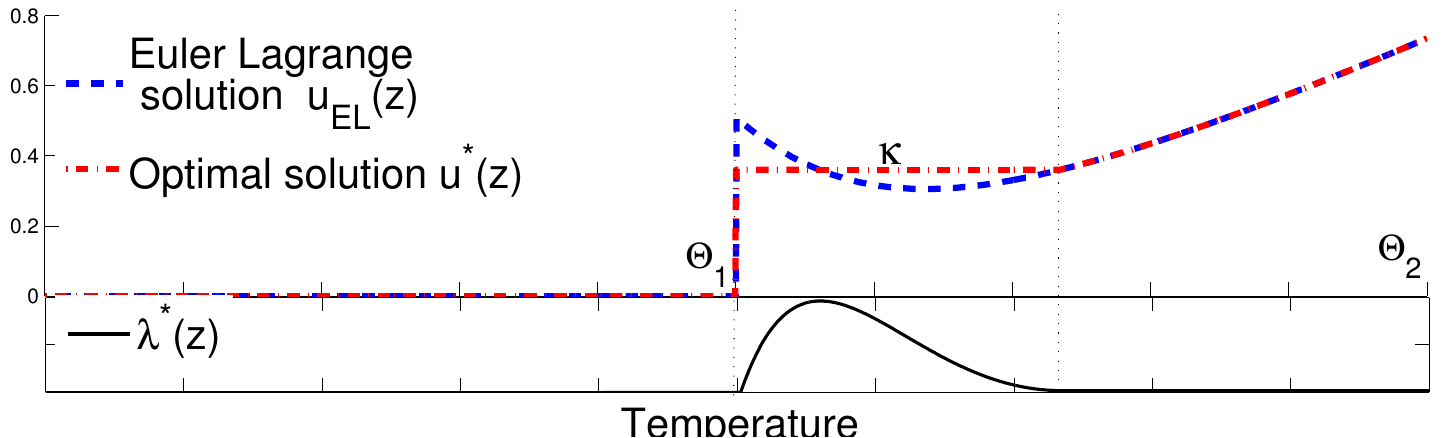}
\caption{The Euler Lagrange from Section \rom{5} and the corresponding  optimal solution obtained from the Pontryagin's minimum principle.}
\label{ustar-fig}
\vspace{-4mm}
\end{figure}

Now we consider  $u_{EL}(z)$ obtained in Section \rom{4}. Since we require $u^*(\Theta_2)=1$, there are jumps at $\Theta_1$ and $\Theta_2$. Next if $u_{EL}(z)$ is decreasing in $[\Theta_1,\theta]$ then $\lambda^*(z)$ is not constant in $[\Theta_1,\theta]$ and $u^*(z)$ is a constant in $[\Theta_1,\theta]$, let $u^*|_{z\in [\Theta_1,\theta]}=\kappa$. Since $\lambda(z)$ can not be negative, $\kappa=u^*(\Theta_1)<u_{EL}(\Theta_1)$ and also since $\lambda(\Theta_2)$ needs to be zero,  $u(\Theta_2)>\min_{z \in [\theta,\Theta_2]}\{u_{EL}(z)\}$, so that $\lambda(z)$ decreases when $u_{EL}(z)<\kappa$. $\lambda(z)$ decreases to zero, and subsequently stays at zero, and $u^*(z)$ remains equal to  $u_{EL}(z)$ till $\Theta_2$. Note that $u^*(z)$ has a discontinuity at $\Theta_2$, since we require $u^*(\Theta_2)=1$. This is illustrated in Figure \ref{ustar-fig}. 

\begin{figure}[!h]
\centering
\includegraphics[height=25mm,width=85mm]{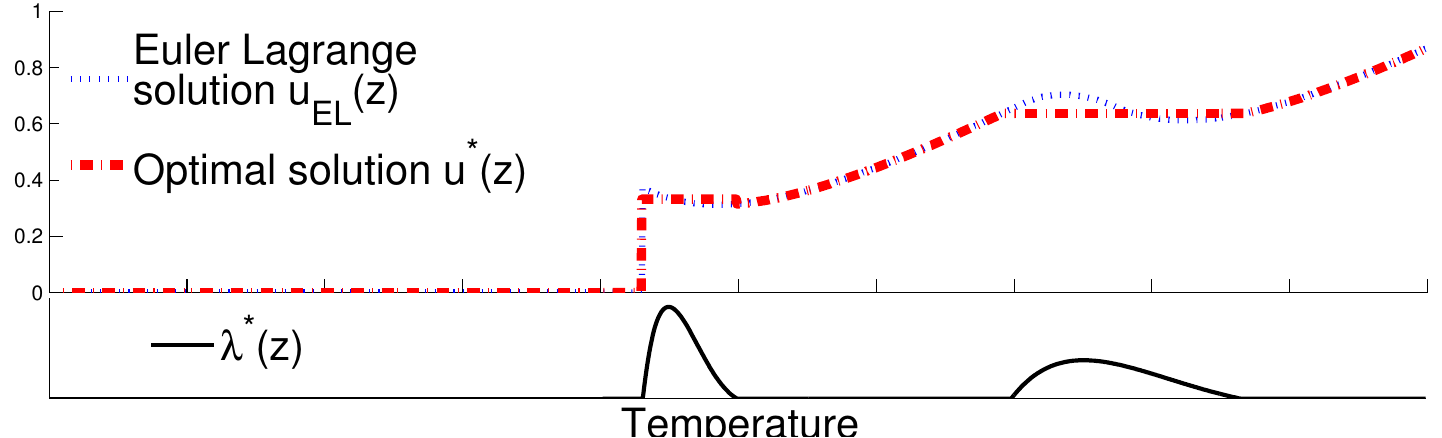}
\caption{ The optimal solution in the general case will remain equal to Euler Lagrange solution in some intervals and will remain constant in other intervals.}
\label{ustar-gen}
\end{figure}

Notice that the above $\kappa$ satisfies $\int_{y_1}^{y_2} (\kappa - u_{EL}(z))w(z)dz=0$, for $y_1:=\sup \{z<\theta:u_{EL}(z)<\kappa \}$ and $y_2:=\inf\{z>\theta:u_{EL}(z)>\kappa\}$. For a general $u_{EL}(z)$, we can find an optimal solution $u^*(z)$ in a similar way. We first find the partition  $\{\Theta_1=\theta_0,\theta_1,...,\theta_N=\Theta_2\}$, such that (wlog)  $u_{EL}$ is increasing in $[\theta_{2i}, \theta_{2i+1}]$, and decreasing in $[\theta_{2i-1},\theta_{2i}]$, and finding $\kappa_i$ such that $\int_{y_{2i-1}}^{y_{2i+i}}(\kappa_i-u_{EL}(z))w(z)dz=0$, for $y_{2i-1}=\sup\{z<\theta_{2i}:u_{EL}(z)<\kappa_i\}$ and $y_{2i+1}=\inf\{z>\theta_{2i}:u_{EL}(z)>\kappa_i\}$. This is shown in  Figure \ref{ustar-gen}.

We denote this transformation of the Euler-Lagrange equation to find a distribution function by the operator $\mathcal{P}[u_{EL}(.)]$, where $\mathcal{P}:\mathbb{R}^{[0,\theta_2]}\to\mathcal{U}$

\subsection{Extensions under multiple comfort level settings}
Consider 3 upper comfort levels $\Theta_1,\Theta_2,\Theta_3$. In this situation, when the comfort range for identical loads toggles to $\Theta_2$, all the loads whose set-points are between $\Theta_1$ and $\Theta_2$ remains unaffected. Therefore, under the event $\{\Theta_M(t)=\Theta_2\}$ the cost will be a function of fraction of set-points below $\Theta_2$, i.e. $u(\Theta_2)$.  
In particular, 
\begin{align*}
\text{Power cost}=&\text{Cost when $\{\Theta_M=\Theta_3\}$}+\\
&\text{Cost when $\{x\geq\Theta_M=\Theta_2\}$}+\\
&\text{Cost when $\{x\geq \Theta_M=\Theta_1\}$}.\\
C_{pow}[u]=\int_0^{\Theta_3}&D^{\Theta_2}(z)(hu(z)+\\
&(h+c)(1-u(z)-u(\Theta_2)))^2+\\
&D^z(z)h^2u^2(z)+D^{\Theta_1}(z)((h+c)-u(z))^2dz,
\end{align*}
where $D^{\Theta_i}(z):=-\frac{\delta_{\Theta_i}^z}{dz}=-\frac{d}{dz}\mathbb{P}(X_z=\Theta_i)$ for $i=1,2$, and $D^z(z):=-\frac{\delta^z_z}{dz}=-\frac{d}{dz}\mathbb{P}(X_z=z)$, and
\begin{align*}
\text{Discomfort cost}&=\int_0^{\Theta_3} \mathbb{E}[(x-\Theta_M(t))^+)]^2 u(z)dz\\
&C_{disc}[u]=\int_0^{\Theta_3} \Phi_1(z)u'(z)+\Phi_2(z)u'(z)dz,
\end{align*}
where $\Phi_i(z)=\int_0^z[(x-\Theta_1)^+]^2(p_{1i}+p_{0i})(z)dz$, for $i=1,2$.
Total cost $C[u]=C_{Pow}[u]+\gamma C_{disc}[u]$.

After simplification the cost function can be re-written as, 
$$J[u]=\int_0^{\Theta_3}(u(z)-u_{EL}(z,u(\Theta_2)))^2w(z)dz+C'(u(\Theta_2))$$
For $u \in \mathcal{U}$, and 
\begin{align*}
u_{EL}&(z,u(\Theta_2))=\\
&\frac{\gamma\Phi'(z)+2c(c+h)(D^{\Theta_1}(z)+D^{\Theta_2}(z)(1-u(\Theta_2)))}{2(h^2D^z(z)+c^2D^{\Theta_1}(z)+c^2D^{\Theta_2}(z))},
\end{align*} 
and 
$C'(u(\Theta_2))=\Phi(\Theta_3)+\int_0^{\Theta_3}(h+c)^2(D_1(z)+D_2(z)(1-u(\Theta_2)))-u_{EL}^2(z,u(\Theta_2))w(z) dz$.\\
The calculation of variation problem is difficult to solve because the functional $J[u]$ also contains point evaluations $u(\Theta_2)$. We propose to solve this iteratively, first by replacing $u(\Theta_2)$ with a variable $v$, and obtaining the optimizer $u^*(z)$ for a fixed $v$,  and then updating $v$ such that $|u^*(\Theta_2)-v|$ decreases in each step.  Thus we can obtain an optimal solution $u^*(z)$ with $v=u^*(\Theta_2)$.

For a initial guess $v_0$ of $u^*(\Theta_2)$ the problem reduces to minimizing $J[u]=\int_0^{\Theta_3} (u-u_{EL}(z,v_0))$, whose solution from Section (\ref{sec:pmp}) is obtained as $\mathcal{P}[u_{EL}(.,v_0)](z)$.
\begin{lem}
let $v$ is such that $v=\mathcal{P}[u_{EL}(.,v)]$, then $v$ lies between $v_0$ and $\mathcal{P}[u_{EL}(.,v_0)](\Theta_2)$.
\end{lem}
\begin{proof}
As $\mathcal{P}[u]$ is increasing in $u$ and $u_{EL}(.,v)$ is decreasing in $v$, if $v_0\leq v$ then $v\leq \mathcal{P}[u_{EL}(.,v)]$, on the other hand, if $v_0\geq v$, then $v\geq \mathcal{P}[u_{EL}(.,v)]$, which yields the desired result. 
\end{proof}

Therefore, it makes sense to define a range in which $v$ lies, let $v_0^\uparrow=\max(v_0,\mathcal{P}(u_{EL}(.,v_0))(\Theta_2))$, and  $v_0^\downarrow=\min(v_0,\mathcal{P}(u_{EL}(.,v_0))(\Theta_2))$. 

For ($n$+1)-th iteration we update the values of $v_n,v_n^\uparrow,v_n^\downarrow$ as, $v_{n+1}=\frac{v_n^\uparrow+v_n^\downarrow}{2}$, $v_{n+1}^\uparrow=\min(v_{n}^\uparrow,\max(v_n,\mathcal{P}[u_{EL}(.,v_n)](\Theta_2)))$, and $v_{n+1}^\downarrow=\max(v_{n}^\downarrow,\min(v_n,\mathcal{P}[u_{EL}(.,v_n)](\Theta_2)))$.

\begin{lem}
The iteration tuples $(v^\downarrow_i, v^\uparrow_i)$ converges to the same unique value as $i\to \infty$. That is, there exist a $v$ s.t. $v^\downarrow_i \to v$, and $v^\uparrow_i \to v$ as $i \to \infty$. Morever such v is a fix-point of the function  $\mathcal{P}(u_{EL}[.,v)]$, i.e. $v$ satisfies  $v=\mathcal{P}(u_{EL}[.,v)]$.
\end{lem}
\begin{proof}
It is clear by induction that for all $i$, $v^\uparrow_i \geq v^\downarrow_i$, since for all $i$, as $v_i=\frac{v_i^\uparrow+v_i^\downarrow}{2}$, we have $\max(v_i,\mathcal{P}[u_{EL}(.,v_i)](\Theta_2))>v_i^\downarrow$ and $\min(v_i,\mathcal{P}[u_{EL}(.,v_i)](\Theta_2))<v_i^\uparrow$. 

Also the sequences $\{v_i^\uparrow\}$ and $\{v_i^\downarrow\}$ are both monotonically decreasing and increasing respectively, therefore they must converge in a bounded set. It remains to show that both of these sequences converges to same point. To show that, we will prove that the range spanned by $v_i^\uparrow$, and $v_i^\downarrow$ decreases with $i$, in particular,  $v_{n+1}^\uparrow-v_{n+1}^\downarrow\leq\frac{v_{n}^\uparrow-v_{n}^\downarrow}{2}$, which concludes the proof, as $v\in [v_i^\downarrow, v_i^\uparrow]$.

Assume, first that $v_i\leq\mathcal{P}[u_{EL}(.,v_i)](\Theta_2)$, then \begin{align*}
v_{i+1}^\uparrow-v_{i+1}^\downarrow & =\min{(v_i^\uparrow, \mathcal{P}[u_{EL}(.,v_i)](\Theta_2))}-v_i\\
& \leq v_i^\uparrow-v_i=\frac{v_i^\uparrow-v_i^\downarrow}{2}.
\end{align*} 
On the other hand, if $v_i>\mathcal{P}[u_{EL}(.,v_i)](\Theta_2)$, then
\begin{align*}
v_{i+1}^\uparrow-v_{i+1}^\downarrow &= v_i-\max{(v_i^\downarrow, \mathcal{P}[u_{EL}(.,v_i)](\Theta_2))} \\
&\leq v_i - v_i^\downarrow = \frac{v_i^\uparrow-v_i^\downarrow}{2}.\\
\end{align*}
That concludes to show that there is a unique point $v$ and $v_i^\uparrow, v_i^\downarrow \to v$ as $i\to \infty$.
\end{proof}

\begin{figure}[!h]
\centering
\includegraphics[height=40mm,width=90mm]{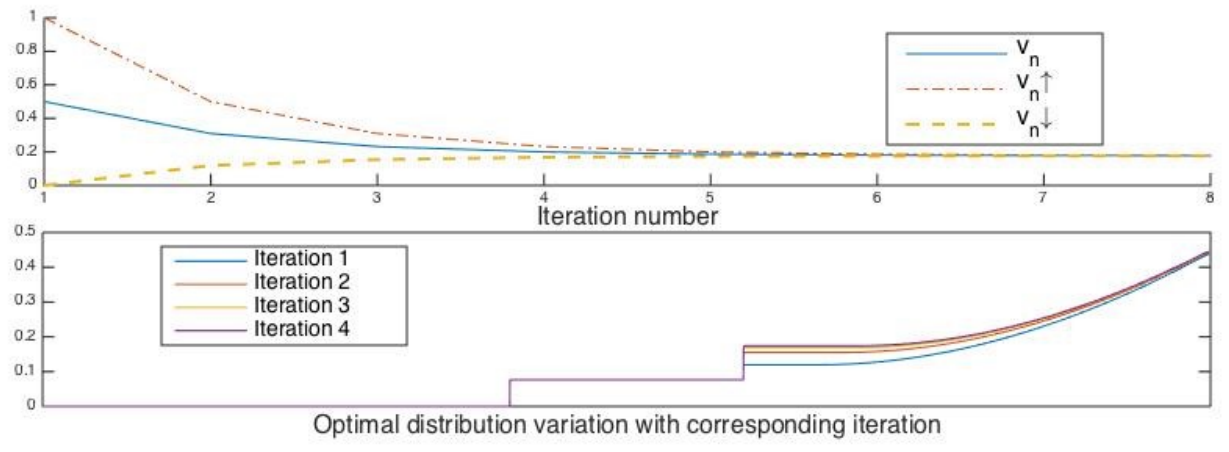}
\caption{Convergence of the parameters $v_n^\uparrow, v_n^\downarrow,$ and $v_n$ with iteration numbers along with the optimal solutions $u^*(.,v_i)$ for $i=1,2,3,4$.}
\label{Conergence}
\end{figure} 
The same technique can be used for more than 3 comfort settings, In particular for $\mathcal{C}$ comfort setting we will have $\mathcal{C}-2$ such tuples $(v^\uparrow,v^\downarrow)$, which can be iterated in a round-robin fashion to obtain the optimal solution. 
\subsection{Extension under multiple wind models}

Consider the case when we have a binary model for comfort setting and, ternary model for wind states, namely 0,1,2. In state 0, the wind is not available so houses heat at the rate $h$. In state 2, all houses can cool at the maximum rate $c$. In addition, we have an intermediate state 1, in which all the houses can be cooled at a rate  $c/2$. 

\begin{figure*}[hb]
\scriptsize
\begin{IEEEeqnarray}{rCh}
u_{EL}(z,\vec{v})=\frac{\gamma \Phi'(z)+2c(c+h)(D^z_{\Theta_1}(z)+\sum_{j=2}^\mathcal{C}(1-v_i)D^z_{\Theta_j}(z))+2c\sum_{i=1}^{\mathcal{W}-1}((1-\frac{i}{\mathcal{W}-1})\hat{D}_{\Theta_1}(z)-\sum_{j=1}{\mathcal{C}}v_i\hat{D}_{\Theta_j}(z)}{2(h^2D^z_z(z)+c^2D^z_{\Theta_1}(z)+\sum_{i=1}^{\mathcal{W}-1}(1-\frac{i}{\mathcal{W}-1})^2c^2\hat{D}(z))}
\label{eq:uelbig}
\end{IEEEeqnarray}
\vspace*{4pt}
\end{figure*}

This system can be solved by using the technique in section (\ref{sec:distr}), to obtain the density functions $p_{00},p_{10},p_{20},p_{10},p_{11},p_{21}$, and the probability masses $\delta_{0,0}, \delta_{0,1},\delta_{\Theta_1,0},\delta_{\Theta_1},\delta_{\Theta_2}$. We notice when $x>\Theta_M$ and wind is in state 1, we can cool the loads at a maximum rate $c$ by providing $c/2$ non-renewable power, which adds to the cost function.

Without going into finite loads case, we directly write the cost function for asymptotic loads scenario. The cost function comes out to be
\begin{align*}
C[u]&=\int_0^{\Theta_2} (-\delta^z_{z})' h^2 u^2(z)\\&
+ (-\delta^z_{\Theta_1,0})'(hu(z)+(h+c)(1-u(z)))^2\\
&+(\hat{D})(z)(\frac{c}{2}(1-u(z)))^2dz + \gamma \int_0^{\Theta_2} \Phi(z)u'(z)dz
\end{align*}

Where $\hat{D}(z)= -\frac{d}{dz}\mathbb{P}(X_z>\Theta_1)$. Denoting $D^z_y=-\frac{d}{dx}\delta_y^z$ for $y=z,(\Theta_1,0), \mbox{ and } (\Theta_1,1)$, and simplifying the above expression to obtain the optimization problem 
$$J[u]= \int_0^{\Theta_2}(u-u_{EL}(z))^2w(z)dz$$
$$s.t. \hspace{5mm} u \in \mathcal{U}$$
$$u(0)=0, u(\Theta_2)=1$$ 
Where $u_{EL}(z)=\frac{\gamma \Phi'(z)+2c(c+h)D^z_{\Theta_1,0}+c\hat{D}(z)}{2(h^2D^z_z(z)+c^2D^z_{\Theta_1}(z)+(\frac{c}{2})^2\hat{D}(z))}$, and $w(z)=h^2D^z_z(z)+c^2D^z_{\Theta_1}(z)+(\frac{c}{2})^2\hat{D}(z)$.

We notice that this is the exact problem as solved in section (\ref{sec:pmp}), whose solution  $\mathcal{P}[u_{EL}(.)](z)$ gives the optimal distribution for set-points. 

The same analysis can be extented for arbitrary $\mathcal{W}>2$ wind states, where the states are numbered such that in state $w\in[0 ,1,...,\mathcal{W}-1]$ the available wind power is sufficient to cool all the houses at a rate $\frac{ic}{\mathcal{W}-1}$. (Therefore the available wind power in state $i$ is $h+\frac{ic}{\mathcal{W}-1}$.)  If there are $\mathcal{C}$ set-points, then the solution of the problem is given by $\mathcal{P}[u_{EL}(.,\vec{v}^*)]$, where $u_{EL}$ is given by equation (\ref{eq:uelbig})

and $\vec{v}^*=(v^*_2,v^*_3,...,v^*_\mathcal{C})$ is the fixed point of the equations $$v^*_j=\mathcal{P}[u_{EL}(.,\vec{v}^*)](\Theta_j)$$ for $j=(2,3,...,\Theta_\mathcal{C})$.

\subsection{Numerical solution for non-homogeneous loads}
The solution for homogeneous loads scenario, with uniform change of set-points is not easy to generalize. In a general case, there can be many sources of in-homogeneity, the cooling and heating rates could be different for different loads, the set-point changes could be arbitrary. All such factors could leads to non-coupled temperature trajectories. Computation of the overall power is difficult under non-coupled temperature case, as the complexity of the state-space increases geometrically in this case. 

We can use numerical simulation to get an estimate of the optimal solution for a general scenario. The idea is to calculate the optimal solution for a population $N$ of loads, which will give us the density function $\tilde{u}(x)=\frac{1}{N}\sum_{i=1}^NH(x-z^*_i)$ as an approximate of 

We used the coupling from the past algorithm to estimate the joint probability distribution $P(\vec{X}<\vec{x})$ from the perfect samples. The cost function comes from the joint distribution as 

\begin{align*}
\tilde{u}(z)=&\sum_{\substack{S\subset\{1,2,...,N\}\\R\subset\{1,2,...,N\}-S}} (\frac{1}{N}\sum_{i\in S, j\in R} \left(P^c_i(x_i)+P^h_j(z_i))\right)^2 \times \\ 
&\hspace{40mm}\mathbb{P}(X_S=z_S\cap X_R>\Theta_R) \\
&+\sum_{i=1}^N \gamma\frac{1}{N} \mathbb{E}[(X_i-\Theta_i)^+]^2
\end{align*}

Where $P_i^h (z_i) :=$ Power required to maintain the temperature of house-i at set-point $z_i $, and $P_i^c(x_i) := $Power required to cool the house-i at temperature $x_i$. 

For simplicity sake, consider the case when the loads are identical to the previous sections. But the set-point changes are independent. The cost function simplifies to

\begin{align}
\tilde{u}(z)=&\sum_{\substack{S\subset\{1,2,...,N\}\\R\subset\{1,2,...,N\}-S}} \left(h\frac{|S|}{N}+(h+c)\frac{|R|}{N}\right)^2 \times \\ &\hspace{30mm}\mathbb{P}(X_S=z_S\cap X_R>\Theta_R) \\
&+\sum_{i=1}^N \gamma\frac{1}{N} \mathbb{E}[(X_i-\Theta_i)^+]^2 \label{eqn:numerical_cost}
\end{align}

Where for any set $U$, we denote the event $X_U>Y_U$ as $\{X_i>Y_i, \forall i \in U\}$.

\begin{figure}[!h]
\centering
\includegraphics[height=25mm,width=85mm]{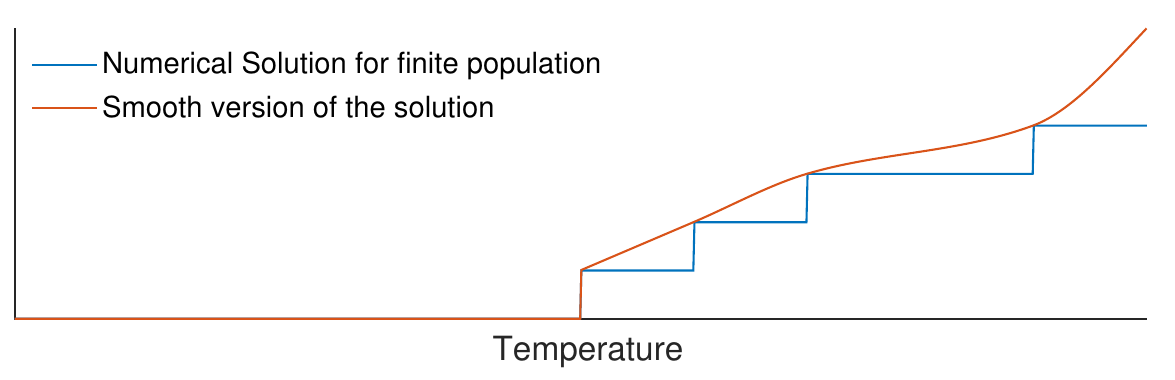}
\caption{ The numerical solution obtained from using Propp's Wilson algorithm, for five loads scenario when the set-point is allowed to change independently}
\label{fig:coulpling}
\end{figure}

The solution $\tilde{u}(\{z_i\})$ resulting from optimization cost function (\ref{eqn:numerical_cost}) may have discontinuities, therefore using this as a distribution to generate set-points will result in a accumulation at the points of discontinuity. One may use interpolation, or smoothing the discrete distribution by any convolution kernal $g(x)$ using operator  $K[u](x)=\int g(x-z)u(z)dz$ to resolve this issue. This is shown in Figure (\ref{fig:coulpling}).

\subsection{Simple heuristic approaches for sub-optimal threshold policy}

In a practical scenario where the air-conditioner parameters, comfort range values, etc., are not known, it is difficult to analyze and propose an optimal threshold policy. A suboptimal heuristic which adaptively updates the thresholds' distribution to minimize the overall cost is desired to tackle this scenario.

A general heuristic method can be proposed when the distribution is chosen from a class $\mathcal{F}_\alpha$ of distributions, where a function $\phi(.,\alpha) \in \mathcal{F}_\alpha$ is characterized by a parameter $\alpha$. A sequence $\{\phi(.,\alpha_n)\}\subset\mathcal{F}_\alpha$ is chosen to achieve $\widehat{J}(\phi(.,\alpha_n)) \geq \widehat{J}(\phi(.,\alpha_{n+1})$. Here $\widehat{J}(\phi)$ is an estimate of the average cost, when $\phi$ is the chosen threshold distribution. The total grid power consumed and the total discomfort cost are assumed to be accessible to the aggregator, without any loss of privacy information from the end-user. The aggregator then estimates the average cost for a long duration and adapts the heuristic distribution to achieve the low overall cost.  

Several classes of distributions $\mathcal{F}_\alpha$ can be chosen. One such choice is the polynomial distribution on $[\Theta_m,\Theta_M]$ i.e.,
\begin{align*}
\mathcal{F}_\alpha^{N}([\Theta_m,\Theta_M]) = \{\sum_{i=0}^N \alpha_iz^i \geq 0: \sum_{i=1}^N i\alpha_iz^{i-1} \geq 0 \\ 
\mbox{ for } z \in [\Theta_m, \Theta_M]\},
\end{align*}

However, as we have shown in the previous sub-sections that there is a discontinuity at the intermediate comfort levels, a polynomial distribution is unable to fit properly.

We therefore consider a set of piecewise polynomials defined on a partition of comfort range, i.e., restriction of the overall distribution on a segment of comfort range to one such polynomial above. The parameterized class of distributions becomes,  
\begin{align*}
\mathcal{F}_\alpha^{(N,T)}([\Theta_m,\Theta_M]) &= \{ (f_1,f_2,..f_{2^T}): \\ 
&\hspace{-5mm} f_i \in \mathcal{F}^N_{\alpha_i}([\Theta_m+\frac{(i-1)\Delta \Theta}{2^T},\Theta_m+\frac{i\Delta \Theta}{2^T}]), \\
 &f_i(\Theta_m+\frac{i\Delta \Theta}{2^T}) \leq f_{i+1}(\Theta_m+\frac{i\Delta \Theta}{2^T}) \},
\end{align*} 

where $\Delta \Theta=\Theta_M-\Theta_m$. For large $N$ the constraint set $\sum_{i=1}^N i\alpha_iz^{i-1} \geq 0$ becomes complex. For simplicity of implementation, assume that $N=0$, i.e. the distribution is piecewise constant in $[\Theta_m, \Theta_M]$, i.e, we would need an adaptive rule for a set of parameters $\{\alpha_i\}_1^{2^T}$, s.t. $\alpha_i\leq \alpha_{i+1}$, and $\alpha_i \geq 0$, and the distribution $f|_{[\Theta_m+\frac{(i-1)\Delta \Theta}{2^T},\Theta_m+\frac{i\Delta \Theta}{2^T}]}\equiv\alpha_i$.

\subsubsection{Algorithm for heuristic policy}
First we assume that the distribution is piecewise constant. In our first approach we suppose that there are fixed partitions (T=constant). The adaptive law is assumed to be similar to gradient descent $(\dot{\alpha_i} = -\epsilon\frac{\partial J}{\partial \alpha_i}$), where $\epsilon$ is the learning rate parameter. In discrete system we have used $\alpha_i(k+1)=\alpha_i(k)-\epsilon\frac{\widehat{J(k)}-\widehat{J(k-1)}}{\alpha_i(k)-\alpha_i(k-1)}$, where $k$ is the time index, and $\widehat{J(k)}$ is the simulated cost at time $k$. To make the algorithm robust, we stop the adaptation when the increase in the cost is smaller than a value $\Delta_J$   

In the successive refining approach, we double the number of partitions each time after finding an (sub)optimal distribution. Each (sub)optimal solution for one partition size is considered as the initial solution for the optimization of the next stage where we double the number of partitions. 

There are several benefits of this approach over the fixed partition one. First, this algorithm reduces the cost
more rapidly than fixed partition scheme, as initially when the partitions are bigger, adaptation results in a bigger change in overall cost reduction. Second, we can stop the sub-partition after the cost reduction is not significant, thus it saves the un-necessary partitioning computation and adaptation. Third, using this approach also reduces search space requirements for adaptation, as the (sub)optimal solution at a coarser level gives a good starting point for the finer level partition.  
\begin{figure}[!h]
\centering
\includegraphics[height=75mm,width=85mm]{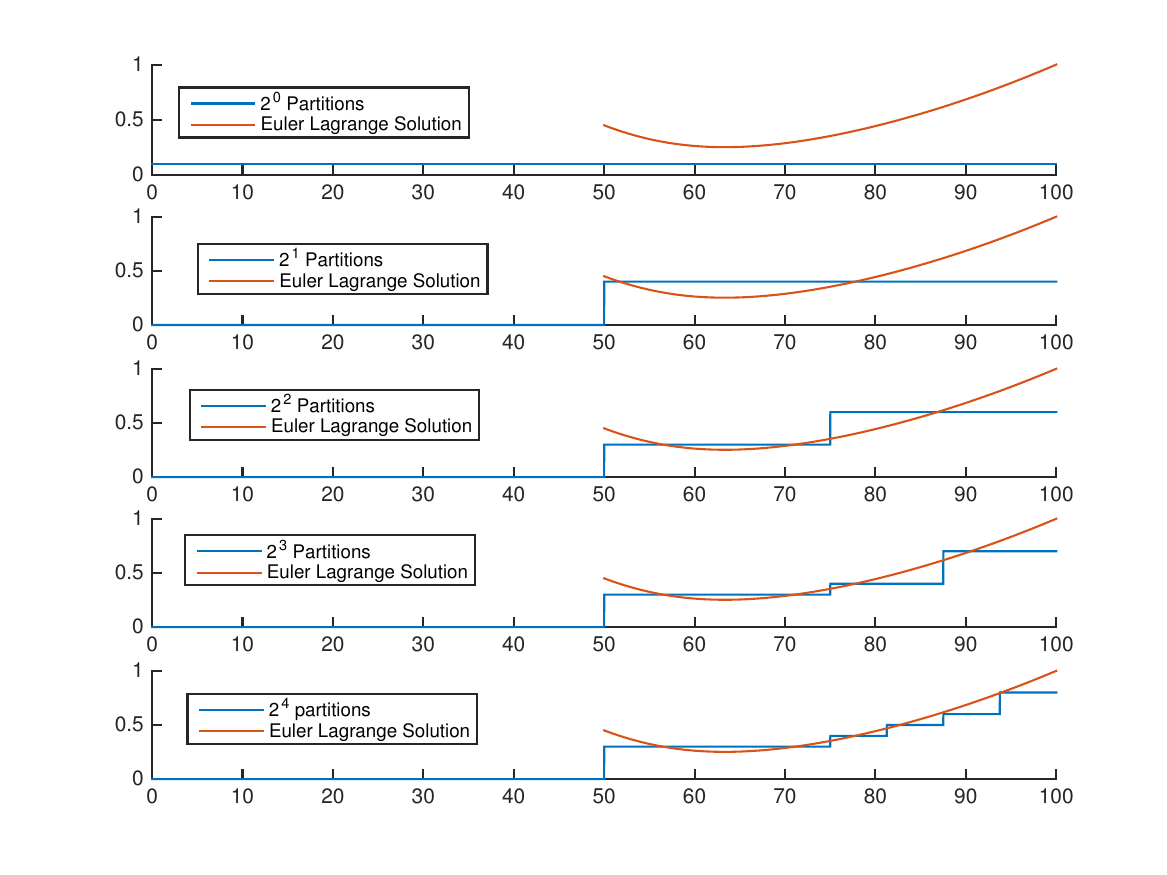}
\label{fig:pw_constt}
\caption{Successive refinement heuristic with piecewise constant distribution, where the distribution is optimized for a fixed partiton size. Subsequently each partition is subdivided into two and further optimization is done with previous optimal solution as the initial guess.}
\end{figure}

Using a piecewise constant distribution has an issue of discontinuity, i.e., the TCL set-point can accumulate the points to discontinuity. To resolve this minor issue, we considered  piecewise linear distribution, where the distribution is the line-segment joining at mid-points of the partitions at levels $\alpha_i$, i.e. the class 
\begin{align*}
&\mathcal{F}_\alpha^T(z)=\\
&\{\frac{z-\alpha_{i-1}}{\alpha_i-\alpha_{i-1}} for z \in [\Theta_m+\frac{\Delta \Theta}{2^T}(i-\frac{1}{2}),\Theta_m+\frac{\Delta \Theta}{2^T} (i+\frac{1}{2})], \\ 
&0=\alpha_0\leq \alpha_1 \leq \alpha_2 ... \leq \alpha_N \leq \alpha_{2^T+1}=1  \},
\end{align*}
where $\Delta \Theta= \Theta_M-\Theta_m$.
\begin{figure}[!h]
\centering
\includegraphics[height=75mm,width=85mm]{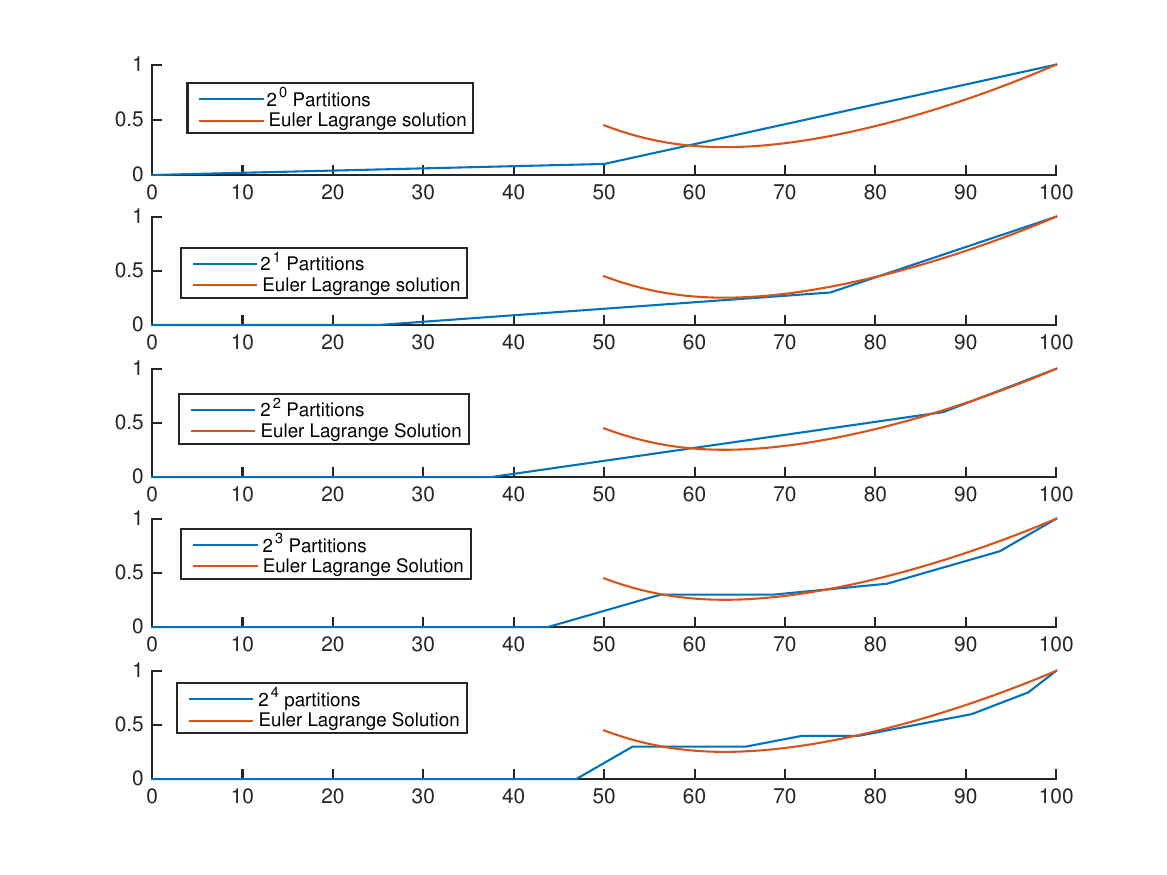}
\label{fig:pw_linear}
\caption{Successive refinement heuristic with piecewise linear distribution.}
\end{figure}
\section{conclusion}
\label{section:conclusion}
We have considered the problem of utilizing an intermittent renewable energy source such as wind to support thermal inertial loads in a microgrid environment. In the first part of this paper we have focused on the issue of reducing non-renewable power variations while adhering to the comfort specifications of end-users of the loads. We have analyzed a scenario where a common wind source is available to support a number of identical loads. We have identified a  key factor of comfort range variation, due to which, even identical loads are treated differently. We have proposed an heuristic method to generalize the nature of the optimal demand response to a large number of loads, for which the exact optimal solution is difficult to obtain. 

In the second part of this paper, we have considered an additional issue of the privacy of the end-user. We have proposed a simple architecture where no information from loads is conveyed, and therefore no privacy is lost. We have calculated the optimal solution for a continuum limit of loads, which can used to control the collective behavior of the loads, without knowing their individual temperatures. We have shown that an explicit solution is analytically computable in a number of scenarios. For the cases where analytical solution is not available but parameters, such as ambient heating rate, cooling rates, are known, we have demonstrated a numerical approach to compute the solution. When such parameters are not available, we proposed and demonstrated an adaptive heuristic approach to obtain reasonable demand response from thermal loads. 

Several directions are open for further research. One direction is to generalize from the microgrid environment to the electric grid, where no information about renewable source is available, and several market structures determine the price of non-renewable consumption and utilization of renewable power sources. Another direction is to incentivize the end user to adhere to a comfort range prescription, where we allow comfort range changes but incentivize (or penalize) the desired (or undesired) comfort range changes.

\bibliographystyle{ieeetr}

%




\section*{Acknowledgment}

\end{document}